\documentclass[aps,preprintnumbers,showpacs,nofootinbib,a4paper]{revtex4}
\usepackage{subfig}
\usepackage{graphicx}% Include figurefiles
\usepackage{dcolumn}
\usepackage{epsfig}
\usepackage{bm}
\newcommand{\eq}{\begin{eqnarray}}
\newcommand{\en}{\end{eqnarray}}

\newcommand{\ba}[1]{\begin{eqnarray} \label{(#1)}}
\newcommand{\ea}{\end{eqnarray}}

\newcommand{\newc}{\newcommand}
\newc{\lra}{\leftrightarrow}
\newc{\beq}{\begin{equation}}
\newc{\eeq}{\end{equation}}
\newc{\barr}{\begin{eqnarray}}
\newc{\earr}{\end{eqnarray}}
  \def\vbf{\mbox{\boldmath $\upsilon$}}
	\def\sbf{\mbox{\boldmath $\sigma$}}
\begin{document}

\topmargin -0.50in
\title {Searching for dark matter axions  via atomic excitations} 

\author{J. D. Vergados}
\affiliation{{\it University of Ioannina,  Ioannina, Gr 451 10,  Greece and Center for Axion  and  Precision  Physics  Research, IBS, Daejeon 34051, Republic of Korea}}
\author{F.T.  Avignone III}
\affiliation{{\it University of South Carolina, Columbia, SC 29208, USA}}
\author{S. Cohen}
\affiliation{{\it University of Ioannina,  Ioannina, Gr 451 10,  Greece }}
\author{R. J. Creswick}
\affiliation{{\it University of South Carolina, Columbia, SC 29208, USA}}
%\email{fedor.simkovic@fmph.uniba.sk}
\begin{abstract}
The possibility of axion detection by observing axion induced atomic excitations as  suggested by Sikivie is discussed. The atom is cooled at low temperature and it is chosen to posses three levels. The first is the ground state, the second is completely empty chosen so  that the energy difference between the two is close  to the axion mass. Under the spin induced axion-electron interaction an electron is excited from the first to the second level. The presence of such an electron there can be confirmed by exciting it further via 
 radiation of a suitably chosen photon energy to the appropriately  selected third level, which is also empty, and lies  at a higher excitation energy.  From the observation of its subsequent de-excitation one infers the presence of the axion.
%From the observation of its subsequent de-excitation one infers the presence of the axion.
% Thus  the presence of the axion can be inferred from the  de-excitation of the second level  to the ground state.
 The system is in a magnetic field so that the energies involved can be suitably adjusted. Since the axion is absorbed by the atom the cross section  exhibits resonance behavior. Using an axion-electron coupling  indicated by the limit obtained by the Borexino experiment,  reasonable axion absorption rates have been obtained for various atomic targets.
\end{abstract}
%\pacs{ 14.60.Pq; 13.15.+g; 23.40.Bw; 29.40.-n; 29.40.Cs}

%\keywords{ neutrino anomaly, sterile neutrino oscillations, neutrino-electron scattering, spherical TPC.}
%\keywords{
% neutrino anomaly, sterile neutrino oscillations, neutrino-electron scattering, spherical TPC.}
%PACS numbers:13.15.+g, 14.60Lm, 14.60Bq, 23.40.-s, 95.55.Vj, 12.15.-y.\\\\\
\date{\today}

\maketitle
{\bf keywords}:\\ Axion detection, axion dark matter, atomic excitations, B induced level splitting, ME of  magnetic moment, \\narrow resonances, laser beams,  frequency and magnetic field scan,  event rate 
%%%%%%%%%%%%%%%%%%%%%%%%%%%%%%%%%%%%%%%%%%%%%%%%%%%%%%%%%%%%%%%%%%%%%
%\include{matrix}
\section{Introduction}
In the standard model there is a source of CP violation from the phase in the Kobayashi-Maskawa mixing matrix. This, however, is not large enough to explain the baryon asymmetry observed in nature.
Another source is the phase in the interaction between gluons ($\theta$-parameter), naively  expected to be of order unity.
The non observation of elementary electron dipole moment limits its value to be $\theta\le10^{-9}$. This has been known as the strong CP problem. 
A solution to this problem has been the P-Q (Peccei-Quinn) mechanism. 
In extensions of the S-M, e.g. two Higgs doublets, the Lagrangian has a global P-Q chiral symmetry $U_{PQ}(1)$, which is spontaneously broken, generating a Goldstone boson, the axion (a). In fact 
the axion has been proposed a long time ago as a solution to the strong CP problem \cite{PecQui77} resulting to a pseudo Goldstone Boson \cite{SWeinberg78,Wilczek78}. The two most widely cited models of invisible axions are the KSVZ (Kim, Shifman, Vainshtein and Zakharov) or hadronic axion models \cite{KSVZKim79},\cite{KSVZShif80}
and the  DFSZ (Dine, Fischler, Srednicki and Zhitnitskij) or GUT axion model \cite{DineFisc81},\cite{DFSZhit80}.  This also led  to the interesting scenario of the  axion being a candidate for dark matter in the universe \cite{AbSik83,DineFisc83,PWW83} and it can be searched for by real
experiments \cite{ExpSetUp11b,Duffy05,ADMX10,IrasGarcia12}. For a  review see, see e.g.,  \cite{Bibber16}.
% The relevant phenomenology has  been reviewed \cite{PhysRep16}.\\
%QCD effects violate the P-Q symmetry and generate a potential $(m_a a)^2 /2 $ for the axion field $a=\theta f_a$ with  axion mass $m_a =(\Lambda^2_{QCD})/f_a $, with  MeV minimum at $\theta=0$ ($\Lambda_{QCD}\approx218$ MeV). Axions can be viable if the SSB (spontaneous symmetry  breaking) scale is large $f_a \geq 100$ GeV.
%Thus the  axion becomes a pseudo-Goldstone boson
%An initial displacement $a_i=\theta _i f_a$  of the axion field causes an oscillation with frequency $\omega= m_a $ and energy density $\rho_D = (\theta f_a m_a)^{2 /2} $
%The production mechanism varies depending on when SSB takes place, in particular whether it takes before or after inflation.

\begin{figure}
\begin{center}
\subfloat[]
{
%\rotatebox{90}{\hspace{-0.0cm} {$\left (\prec{\vec \upsilon} .{\hat n}\succ/\upsilon_0\right )_{ang}$}}
%\rotatebox{90}{\hspace{-0.0cm} {$|g_{a\gamma\gamma}|\rightarrow$GeV$^{-1}$}}
\includegraphics[width=0.4\textwidth, height=0.3\textwidth]{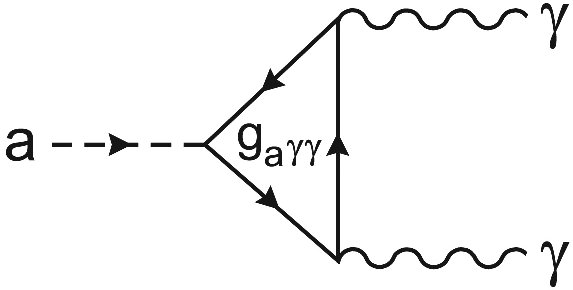}
%{\hspace{-2.0cm} {$m_{a}\longrightarrow\,\mu$eV}}\\
}
\subfloat[]
{
%\rotatebox{90}{\hspace{-0.0cm} {$\left (\prec{\vec \upsilon} .{\hat n}\succ/\upsilon_0\right )_{ang}$}}
%\rotatebox{90}{\hspace{-0.0cm} {$|g_{a\gamma\gamma}|\rightarrow$GeV$^{-1}$}}
\includegraphics[width=0.4\textwidth, height=0.4\textwidth]{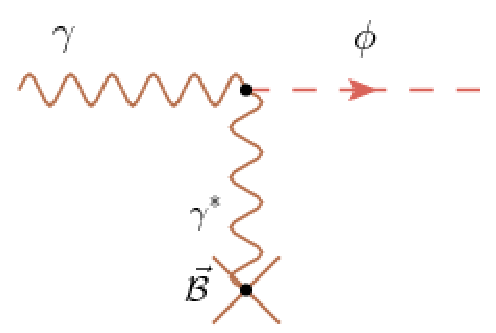}
%{\hspace{-2.0cm} {$m_{a}\longrightarrow\,\mu$eV}}\\
}\\
\subfloat[]
{
%\rotatebox{90}{\hspace{-0.0cm} {$\left (\prec{\vec \upsilon} .{\hat n}\succ/\upsilon_0\right )_{ang}$}}
%\rotatebox{90}{\hspace{-0.0cm} {$|g_{a\gamma\gamma}|\rightarrow$GeV$^{-1}$}}
\includegraphics[width=1.0\textwidth, height=0.5\textwidth]{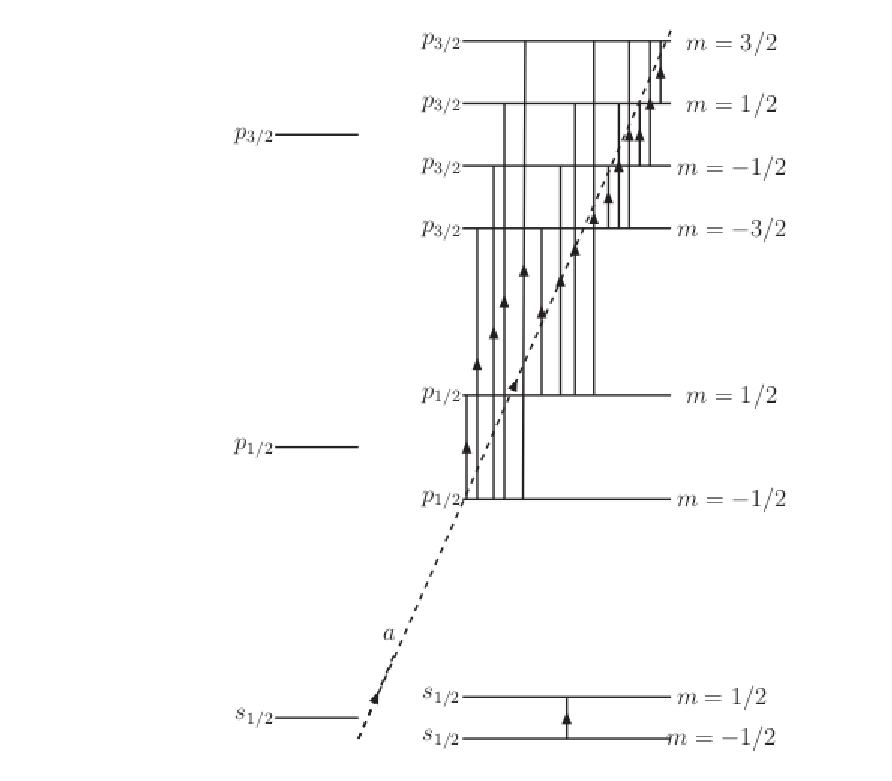}
%{\hspace{-2.0cm} {$m_{a}\longrightarrow\,\mu$eV}}\\
}
 \caption{(a) The  axion to photon interaction, axion photon coupling. (b) The  axion to photon conversion  in the presence of a magnetic field, the Primakoff effect. (c) The axion absorption by an atom at low temperature, via the axion electron spin induced interaction. On the left the spectrum is as it appears  in the absence of a magnetic field. On the right the energy splittings depend on the spin-orbit interaction and  the magnitude of  the magnetic field (the levels are not on scale).  An electron  is moved from an occupied initial level (anywhere the arrows begin) with energy $E_i$ and angular momentum $n\ell j_i$ corresponding the substate $m_i=-j_i$ to a level with energy $E_f$ with an angular momentum $n\ell j_f$ , which may be anywhere the arrows provided that $m_f=-j_i+0,\pm 1$ consistent with the angular momentum selection rules. The picture is drawn for one electron configurations, but it can be generalized to multi-electron configurations. In all cases only states with the same radial and orbital  quantum numbers can be connected via the spin operator. When the total angular momentum of the two states is the same, the splitting is small and  due to the magnetic field.}
 \label{AxionPhoton}
  \end{center}
  \end{figure}

% The axion field is homogeneous over a large de Broglie wavelength, oscillating in a coherent way, which makes it  ideal cold dark matter candidate . 
It has  been recognized long time ago that the axion is an ideal cold dark matter candidate, especially  in the mass range $10^{-6}$ eV$\leq m_a \leq 10^{-3}$ eV, by Sikivie \cite{Sikivie83}, and others, see, e.g, \cite{PriSecSad88}.
%Since the  axions are extremely light and non relativistic, it is impossible to detect axions as dark matter particles in a traditional way, i.e.  via scattering them off targets. 
Thus the  popular experiments  hope to   detect them  by their conversion to photons in the presence of a magnetic field (Primakoff effect), see Fig. \ref{AxionPhoton}(a),(b). The produced photons are detected in a resonance cavity  as suggested by Sikivie \cite{Sikivie83}. In the case of the axion absorption by atoms, see  Fig. \ref{AxionPhoton}(c), the detection can be achieved by directly measuring the photons following the atom de-excitation or as described in the text.

% It is interesting  that recently it has been shown that the axions can be treated in terms of a generic low mass
%non-relativistic scalar field theory  with understood
%equilibrium behavior \cite{GHP-W14} and exploitable
%galactic behavior. As a result  axions are cold non barionic  particles, which, in principle, can be distinguished from other dark matter candidates \cite{Sikivie11}. 
 In fact various experiments\footnote{Heavier axions with larger mass in the 1eV region produced thermally ( such as via the  $a\pi\pi\pi$ mechanism), e.g. in the sun, are also interesting and are searched by CERN Axion Solar Telescope (CAST) \cite{CAST11}.
Other axion like particles (ALPs), with broken symmetries not connected to QCD, and dark photons form  dark matter candidates called WISPs (Weakly Interacting Slim Particles) , see, e.g.,\cite{ADDKM-R10}, are also being searched.}
 such as ADMX and ADMX-HF collaborations 
\cite{ExpSetUp11b},\cite{ADMX10},\cite{Stern14,MultIBSExp}  are  planned and ongoing to search for them.  In addition, the newly established center for axion and physics research (CAPP)  has started an ambitious axion dark matter research program \cite{CAPP}, using SQUID and HFET technologies \cite{ExpSetUp11a}. Their strategy is to run several experiments in parallel to explore a wide range of axion masses with sensitivities better than the QCD axion models \cite{Semer20a},\cite{Semer20b},\cite{SemerArch19}. 
% is also  planning to search for axions.
  \\The allowed parameter space has been presented in a nice slide by Raffelt \cite{MultIBSTh} in the  Multidark-IBS workshop and, focusing on the axion as dark matter candidate, by Stern \cite{Stern14}, derived from Fig. 3 of ref \cite{Stern14}).
  
  Recently some exclusion on the axion masses have been obtained by the ADMX experiments in the range of 2.66-2.81 $\mu$eV \cite{ADMX18}, 2.81-3.39 $\mu$eV \cite{ADMX20} and 3.3-4.2 $\mu$eV \cite{ADMX21} leading to the
  exclusion of a wide   range of axion-photon coupling values predicted in benchmark models of the invisible
  axion, which  solve the strong CP problem of quantum chromodynamics.
	
Since, however, the mass of the axion is not known, it is important to consider other processes for its detection, which may be accessible to a wider window of axion mass. Such may involve, e.g., axion detection via atomic excitations \cite{ZioutSemer88},
\cite{TFSSB18a}, \cite{FSTB18b}. 
%Such experiments in one way or another involve the  photon-to-axion conversion probability in the presence of a magnetic field and, in some cases, exploiting the axion-photon interference  pattern\cite{TFSSB18a}.

In this paper we are going to discuss the possibility of axion detection by observing directly  axion induced atomic excitations, measuring the photons produced in the de-excitation to the ground state,  with sensitivity  to axion masses in the range $10^{-5}$ eV close to 1 eV. Another procedure,  suggested by Sikivie \cite{Sikivie14},  involves an atom  cooled at low temperature, which utilizes three energy  levels. The first is  the ground state. The second  is completely empty, chosen  such that the energy difference between the two is close  to the axion mass. Under the spin induced axion-electron interaction an electron is excited from the first to the second  level. The presence of such an electron there can be confirmed by exciting it further via radiation of suitably chosen photon energy to a third level, which is also empty, and lies  at higher excitation energy. From the observation of its subsequent de-excitation one infers the presence of the axion (see Fig.  \ref{AxionPhoton}c).\\ 
%In addition the presence of the axion can be inferred from the de-excitation of the second level to the ground state.
The  magnetic field employed is  used to split the magnetic m-substates so that the transition energies involved can be suitably adjusted. Furthermore, by suitably adjusting in size, it can determine a window of axion masses to be searched in given  experiment

A crucial parameter in the axion induced atomic excitations is the axion   electron  coupling $g_{ea}/f_a$. The dimensionless quantity has been studied in axion models $g_{ea}$. The quantity $f_a$ with dimension of mass is not known, but  it is believed to be  inversely proportional to the axion mass. In the present calculation  we are going to adopt a value $g_{ea}/f_a$ which coincides with the stringent limit obtained in the Borexino experiment \cite {Borexino12}. We will see that this value leads to a very small cross section. Fortunately,  we will find  that experiments involving dark matter axions are not doomed to be un-observable, since the axion number density in our vicinity of the  galaxy is quite large, owing to the small axion mass.

Our paper will be organized as follows: In section \ref{sec:axionabsorption} we will derive expressions yielding the  rates  for axion absorption  by atoms, in section 	\ref{sec:caxionelec}  we will discuss he axion electron coupling and the range of the axion masses obtained from Borexino limit in conjunction with reasonable axion model parameters $g_{ea}$, in section \ref{sec:axionwidth} we will study the obtained axion widths, in section \ref{sec:AtomicPhys} we will summarize the needed atomic physics input, in section \ref{ec:lowtemp} we will consider the low temperature requirements for the success of the experiments, in section \ref{sec:Rates} we will  present our results for  the expected rates and in section \ref{conclusions} we will summarize our conclusions.

\section{Expressions for rates for axion absorption  by atoms}
\label{sec:axionabsorption}
 We remind the reader that the axion, $a$, is a pseudoscalar particle and its coupling to the electron can be described by a Lagrangian of the form:
\beq
{\cal L}=  \frac{g_e}{f_a}i\partial_{\mu}a\bar{\psi}({\bf p}',s)\gamma^{\mu}\gamma_5\psi({\bf p}, s)
\eeq
where $g_e$ is a coupling constant and $f_a$ a scale parameter with the dimension of energy. For an axion with mass $m_{a}$ it easy to show that  in the non relativistic limit:
\begin{itemize}
\item the time component $\mu=0$ is given by:
\beq
{\cal L}=\langle \phi|\Omega|\phi \rangle,\,\Omega=\frac{g_e m_a}{2 f_a}\frac{\sbf.{\bf q}}{m_e},\,{\bf q=p'-p}
\label{Eq:AxionelInt}
\eeq 
which is negligible for $m_a<< m_e$.
\item The space component, $\mu\ne 0$,
\beq
{\cal L}_{aee}=\langle \phi|\Omega|\phi \rangle,\,\Omega=\frac{g_e}{2 f_a}\sbf.{\bf q},\,{\bf q=p'-p}
\eeq
where ${\bf p}$ and ${\bf p}'$ are the initial and final electron momenta,  $f_a$  the axion decay constant and $\sbf$ the spin of the electron.\\
\end{itemize}
An  interaction of the form of Eq. (\ref{Eq:AxionelInt}) has  been proposed by Sikivie
% as a way of detecting the axion by causing atomic excitations 
% Thus the axion coupling to electrons in the non relativistic limit  has the above form 
  \cite{Sikivie14}  as a way of detecting the axion by causing atomic excitations,
%\beq
%{\cal L}_{aee}=\frac{g_e}{2 f_a}{\bf q}.\sbf
%\eeq
%where $f_a$ is the axion decay constant, ${\bf q}$ the axion momentum and $\sbf$ the spin of the electron.
$g_{a e}$ is the relevant coupling constant to be determined by experiment.\\
The target is selected so that there exist two levels, say $|J_1,m_1\rangle$ and   $|J_2,m_2\rangle$ which result from the splitting of the atomic levels by the magnetic field and they are characterized by the same $n$ and $\ell$ so that they can be connected by the spin operator. The lower one $|J_1,m_1\rangle$ is occupied by electrons but the higher one $|J_2,m_2\rangle$ is completely empty at sufficiently low temperature. It can be populated only by exciting an electron to it from the lower one by the axion field. The occurrence of such an excitation is monitored by a tuned laser which excites such an electron from    $|J_2,m_2\rangle$  to a higher state  $|J_3,m_3\rangle$, which cannot be reached  in any other way, by observing its subsequent decay.

In the present case we are interested in the case that the states $|J_1,m_1\rangle$ and   $|J_2,m_2\rangle$ can be connected via the spin operator, i.e. the two states  must have the same orbital structure and spins that can be reached by the spin operator ( for single particle states, $J_1=j_1$, $J_2=j_2$, they must have  the same $n$ and $\ell$ quantum numbers). We distinguish to cases:\\ i) $J_2=J_1$. In this case the splitting is due to  the magnetic field  yielding about $10^{-4}$ eV/T. This is appropriate for detection of axions around this splitting for the chosen magnetic field.
% in the mass region $10^{-6}-10^{-3}$ eV.
\\ ii) $J_2\ne J_1$, i.e. they correspond to the two spin orbit partners. Then the energy splitting could be in the  eV range and, thus, this arrangement is suitable for the detection of axions with mass in the same range.

 Let us for simplicity assume a single particle  transition. The relevant  matrix element for the transition $j_1,m_1\rightarrow j_2,m_2  $takes the form

\beq 
\langle n \ell j_2  m_ 2 |{\bf q}.\sbf|n \ell j_1  m_ 1\rangle=C_{\ell,j_1,m_1,J_2,m_2 }q_{m_1-m_2}I_{n \ell}({\bf q})
\eeq
	where $C{\ell,j_1,m_1,j_2,m_2 }$ depends on the atomic levels \cite{Vergdos20} and it will be given below, see section \ref{sec:AtomicPhys}, and
	$ I_{n \ell}({\bf q})$ is given by
\beq
 I_{n \ell}({\bf q})=\int d^3{\bf p} \phi_{n \ell}({\bf p+q})\phi_{n \ell}({\bf p})
 \eeq
Since the momentum transfer ${\bf q}$ is small  $ I_{n \ell}({\bf q})\approx 1$. It can be shown that a similar result holds in the case of multi-particle configurations  So the matrix element becomes
\beq
|\mbox{ME}({\bf q})|^2=\left (\frac{g_e}{2 f_a}\right )^2 \left (C_{\ell,J_1,m_1,J_2,m_2 }\right )^2\left(\delta_{m_1,m_2}q^2_0+\frac{1}{2} (q_1^2+q_2^2)(1-\delta_{m_1,m_2})\right )
\eeq
where ${\bf q} $ is the momentum transfer to the atom with $q_0$ its  component  in the direction of the axis of quantization and $q_1$, $q_2$ along the other two  axes.

The cross section becomes
\beq
\sigma=\frac{1}{\upsilon}\frac{1}{2 m_a} |\mbox{ME}({\bf q})|^2 \int \int \frac{d^3{\bf p}_A}{(2 \pi )^3}(2 \pi )^3\delta({\bf q-p}_A) 2\pi \delta (m_a+\frac{q^2}{2 m_a}+E_i-E_f)
\eeq
where ${\bf p}_A$ the momentum transfer to the atom. $ 2 m_a$ is the usual normalization for a boson field, In the above expression we have neglected the tiny recoiling energy of the atom. Thus
\beq
\sigma=\frac{1}{\upsilon}\frac{1}{2 m_a}\left (\frac{g_e}{2 f_a}\right )^2 \left (C_{\ell,j_1,m_1,j_2,m_2 }\right )^2\left(\delta_{m_1,m_2}q^2_0+\frac{1}{2} (q_1^2+q_2^2)(1-\delta_{m_1,m_2})\right )2\pi \delta (m_a+\frac{q^2}{2 m_a}+E_i-E_f)
\eeq

We will now fold the cross section with the axion velocity distribution, assuming  that  with respect to the galactic center is of the Maxwell-Boltzmann type:
\beq
f_g(\upsilon')=\frac{1}{\upsilon_0^3}\frac{1}{\pi \sqrt{\pi}}e^{-\left( \frac{\upsilon'}{\upsilon_0} \right )^2}
\eeq
In the local frame, ignoring for the moment the motion of the Earth, we have $\vbf'\rightarrow \vbf+\upsilon_0 \hat{z}$
\beq
f_{\ell}(\vbf)=\frac{1}{\upsilon_0^3}\frac{1}{\pi \sqrt{\pi}}e^{-\left( y^2+2y \xi+1\right )},\, y=\frac{\upsilon}{\upsilon_0}
\eeq
The integration over the velocity distribution we find:
\beq
\langle y \sigma\rangle=\frac{1}{2 m_a}\left (\frac{g_e}{2 f_a}\right )^2 \left (C_{\ell,j_1,m_1,j_2,m_2 }\right) \Lambda
\eeq
where
\barr
\Lambda&=& \frac{1}{\upsilon_0}\int y f(\vbf) d^3 \vbf\left(\delta_{m_1,m_2}q^2_0+\frac{1}{2} (q_1^2+q_2^2)(1-\delta_{m_1,m_2})\right )2\pi \delta (m_a(1+\frac{1}{2} \upsilon^2)+E_i-E_f)\nonumber\\
&=&\int y dy y^2  (m_a \upsilon_0 y)^2 2\pi \delta (m_a(1+\frac{1}{2} \upsilon_0^2 y^2)+E_i-E_f)J\frac{1}{(\sqrt{\pi})^3},\nonumber\\J_1&=&\int  d \Omega e^{-\left( y^2+2y \xi+1\right )}\left(\delta_{m_1,m_2}\xi^2+\frac{1}{2} (1-\xi^2)(1-\delta_{m_1,m_2})\right )
\label{Eq:deltaint}
\earr
 So  the integration over the angles  yields:
\begin{itemize}
\item In the galactic frame
\beq
J_1=2 \pi J,\,J=e^{-y^2} 2 \pi\int d\xi  \left(\delta_{m_1,m_2}\xi^2+\frac{1}{2} (1-\xi^2)(1-\delta_{m_1,m_2})\right )=\frac{4 \pi}{3} e^{-y^2}
\eeq
which is symmetric.
\item in the local frame we get:
\beq
J_1=2\pi J,\,J=e^{-1-y^2} 2 \pi\int d\xi e^{-2 y \xi} \left(\delta_{m_1,m_2}\xi^2+\frac{1}{2} (1-\xi^2)(1-\delta_{m_1,m_2})\right )
\label{Eq:localJ}
\eeq
that is 
\beq
J=   e^{-1-y^2}\left (\delta_{m_1,m_2}\frac{\left(2 y^2+1\right) \sinh \left(2
   y\right)-2 y \cosh2
   y}{2 y^3}+\left(1-\delta_{m_1,m_2}\right ) \frac{2 y \cosh 2 y-\sinh 2
   y}{4 y^3}\right )
\label{Eq:Jy}
	\eeq
\end{itemize}
Th integration over the magnitude of the velocity is trivial due to the $\delta$ function appearing in Eq. (\ref{Eq:deltaint}). We thus get in the local frame:
\beq
\Lambda=4\sqrt{\pi} m_a  \frac{1}{\upsilon_0}  F_{m_1,m_2}(X), F_{m_1,m_2}(X)=\begin{array}{cc} \frac{1}{2} X e^{-X^2-1}
	\left(\left(2 X^2+1\right)
	\sinh (2 X)-2 X \cosh (2
	X)\right),&m_1=m_2\\\frac{1}{4} Xe^{-X^2-1} (2 X
	\cosh (2 X)-\sinh (2 X)),&m_1\ne m_2\\\end{array}
\label{Eq:Lambda}
\eeq
The extra factor of $X^4$ in going from Eq. (\ref{Eq:Jy}) to Eq. (\ref{Eq:Lambda}) is the result of the integration over the velocity.

Sometimes we prefer to normalize the function $F_{m_1,m_2}(X)$. Then we write
\beq
\Lambda=4\sqrt{\pi} m_a \frac{1}{\upsilon_0}  N_{m_1,m_2}  F^N_{m_1,m_2}(X), N_{m_1,m_2}=\begin{array}{cc} \frac{1}{2} \sqrt{\pi }
	\text{erf}(1),&m_1=m_2\\\frac{e \sqrt{\pi }
		\text{erf}(1)+2}{8 e} ,&m_1\ne m_2\\\end{array}
	\label{Eq:LambdaN}
\eeq
In Eq. (\ref{Eq:LambdaN}) $F^N_{m_1,m_2}(X)$  is a function of $X$ given by
\barr 
F^N_{m_1,m_2}(X)&=&\left \{\begin{array}{cc} \frac{Xe^{-X^2-1} \left(\left(2
		X^2+1\right) \sinh (2 X)-2 X
		\cosh (2 X)\right)}{\sqrt{\pi
		} \text{erf}(1)},&m_1=m_2\\\frac{2X e^{-X^2} (2 X \cosh (2
	X)-\sinh (2 X))}{e \sqrt{\pi
} \text{erf}(1)+2} ,&m_1\ne m_2 \\ \end{array} \right .
\label{Eq:F(X)}
	\earr
with $X$ given by:
\beq
X=\frac{c}{\upsilon_0}\left( \sqrt{2\left(\frac{E_f-E_i}{m_a c^2}-1\right )}\right).
\label{Eq:X(Delta)}
\eeq
% The normalization factor $N_{m_1,m_2}$, introduced to normalize  $F_{m_1,m_2}(X)$, takes the values   $\approx$0.75 for $m_1=m_2$ and $\approx$ 0.28 for $m_1\ne m_2$.
 
 Thus one obtains:
 \beq
 \langle y \sigma\rangle=\frac{1}{2}\frac{1}{\upsilon_0} \left (\frac{g_e}{ f_a}\right )^2 4 \sqrt{\pi}  \left (C_{\ell,j_1,m_1,j_2,m_2 }\right)^2     F_{m_1,m_2}(X).
 \eeq

 The event associated with a flux of particles with velocity $\upsilon$ (per atom in the target) is given by:
 \beq
 R=\Phi_a \sigma
 \label{Eq:Phia}
 \eeq
 where  $\Phi_a$ is the axion flux given by $\Phi_a=\frac{\rho_a}{m_a}\upsilon_0 $ with $\rho_a$ the axion  matter density in our vicinity of the galaxy.   In this work we  will assume that all dark matter in our vicinity is composed of axions. So it is  obtained from the rotation curves and employed in standard dark matter searches,i.e.  $\rho_a=0.3\mbox{Gev/cm}^3$. This leads to a large axion particle density $\frac{\rho_a}{m_a}$ due to the smallness of the axion mass. 

Thus averaging over the velocity distribution we get per atom
 \beq
 \langle R \rangle =\frac{\rho_a}{m_a}\upsilon_0\langle y  \sigma \rangle
 \eeq
  Thus Eq.  (\ref{Eq:Phia}) for N atoms in the target becomes 
 \beq 
 R=R_0(m_a)  \left (C_{\ell,j_1,m_1,j_2,m_2 }\right)^2     F_{m_1,m_2}(X),R_0(m_a)= \Phi_0(m_a) \sigma_0,\Phi_0(m_a)= N \frac{\rho_a}{m_a}\upsilon_0, \sigma_0=2\sqrt{\pi}  \frac{1}{\upsilon_0}\frac{g^2_{ae}}{ f_a^2}
  \label{Eq:rate}
 \eeq
 %In the above expression we have separated the coupling constant $g_e$, which can be determined from experiment.
  $R_0(m_a)$ is written as a product of two constants,  one with the dimension of the flux, which varies inversely proportional to the axion mass and   the other yields the scale of the cross section.
  \section{The axion electron coupling}
  \label{sec:caxionelec}
  A crucial parameter in the present work is $\frac{g_{ae}}{f_a}$. In the past this parameter was derived from existing axion models. In a recent paper 
  	\cite {Borexino12} 
  %	\footnote{G. Bellini et al. (The Borexino Collaboration), Phys. Rev. D 85 , 092003 (2012), arXiv:1203.6258 [hep-ex℄].}
  	 the following limits were obtained:
  	\beq
  	|g_ {Ae} \times m _A | \le 2.0 \times 10 ^{-5} \mbox{eV},\,|g_ {Ae} \times g_{3AN} | \le 5.5 \times 10 ^{-13} \mbox{eV}.
  	\eeq
  	These can be  interpreted  to be the axion electron and the isovector axion nucleon coupling, which in   our notation  are written:
  	\beq
  	|g_ {ae} \times m _a | \le 2.0 \times 10 ^{-5} \mbox{eV},\,| g^3_{aN} | \le 2.8 \times 10 ^{-8} \mbox{eV}
  	\label{Eq:aeaN}
  	\eeq
  	Using now the equation \cite{CorGhVill18}, \cite{VerDivEj22}
  	\beq
  	m_a f _a \approx 6000 \mbox{MeV}^2 
  	\eeq
  	we obtain
  	\beq
  	\frac{g_{ae}}{ f _a }\leq  3.3\times 10 ^{-12}  \mbox{GeV}^{-1},\,\frac{g^3_{aN}}{ f _a }\leq  4.7\times 10 ^{-15}  \mbox{GeV}^{-1}
  	\label{Eq:faeaN}
  	\eeq
  	These couplings are indeed very small. This perhaps explains why for  the Borexino 5.5.MeV solar axion flux on Earth, resulting from the second relation of the equation, is very small. This  is  the reason why, in our recently published  paper \cite{VerDivEj22}
  	% \footnote{John D. Vergados ,  Paraskevi C. Divari  and Hiroyasu Ejiri. Advances in High Energy Physics
  	%	Volume 2022, Article ID 7373365, 
  	%	https://doi.org/10.1155/2022/7373365}, 
  	 we had  to  admit that the nuclear excitations were not detectable. This, of course did not affect  the atomic excitations discussed in the same paper, since the axion  flux for the 14.4 keV solar axions, used in this paper, happened to be much larger. Anyway in the present calculation we consider dark matter axions with flux of a different origin.
  	 
  	 The coupling  $g_{ae}$ is not known. but it has been  investigated \cite{CHVV16, RingSaika15}, in particular in  the context of the DFSZ axion models \cite{DineFisc83, DFSZhit80}.
  	 This leads to:
  	 \beq
  	 g_e=\frac{1}{3}\left (1-\frac{\tan^2{\beta}}{1+\tan \tan^2{\beta}}\right ), \tan{\beta}=\frac{\upsilon_2}{\upsilon_2} \mbox{ or } 
  	 g_e=\frac{1}{3}\cos^2{\beta}
  	 \eeq
  	 Where $\tan{\beta}$ is the ratio of the vacuum expectation values of the two doublets of the model, the parameter $\beta$  is not known, but some sort of theoretical limits exist \cite{Srednicki85}, e.g.
  	 $ \frac{1}{6}<g_e<\frac{1}{3}$.
  	% We will  conservativel the  value $g_e=\frac{1}{6}$
  	 
\begin{figure}
  	 	\begin{center}
  	 		\subfloat[]
  	 		{
  	 			\rotatebox{90}{\hspace{-0.0cm} $m_a\rightarrow$ eV}
  	 			\includegraphics[width=0.3\textwidth]{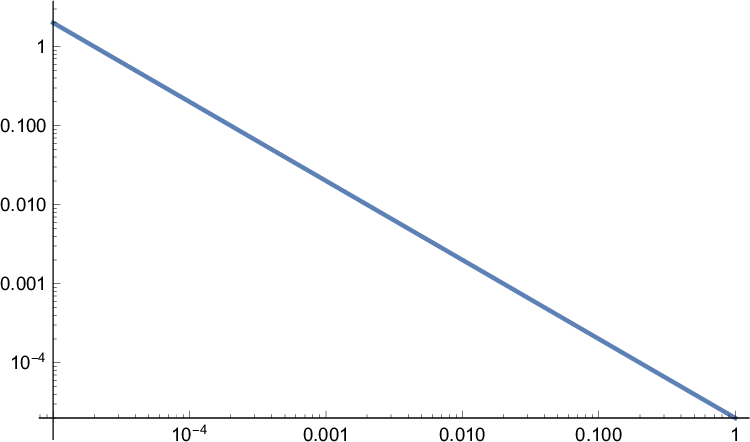}
  	 			%			{\hspace{-3.0cm}$g_{ae} \rightarrow$  }	
  	 		}
  	 		\subfloat[]
  	 		{
  	 			\rotatebox{90}{\hspace{-0.0cm} $\beta=\frac{v_2}{v_1}\rightarrow$ }	
  	 			\includegraphics[width=0.3\textwidth]{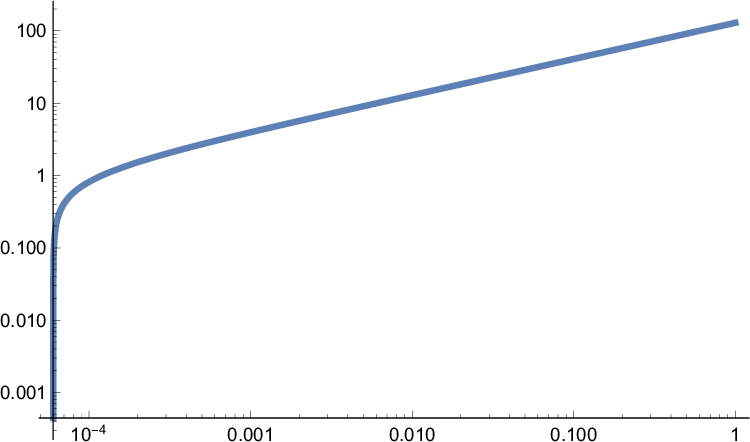}	
  	 			%	{\hspace{-3.0cm}$m_a \rightarrow$ eV }
  	 		}\\
  	 		{\hspace{1.0cm}$g_{ae} \rightarrow$  }		{\hspace{7.0cm}$m_a \rightarrow$ eV }		
  	 		\caption{(a)The axion mass $m_a$ allowed by the coupling given in Eq. (\ref{Eq:aeaN}) as a function of $g_{ae}$. This covers the whole range of $m_a$ of interest in this work. (b) The ratio $\beta=\frac{v_2}{v_1}$ of the   expectation vulues of the Higgs doublets of the DFSZ axion model discussed in the text. as a function of the axion mass $m_a$. }
  	 		\label{fig:avsgae}
  	 	\end{center}
  	 \end{figure}

  	 In the present work  we will assume that the upper limit of Eq.  (\ref{Eq:aeaN}) corresponds to the actual value of $ g_{ae}m_a$. Then, since the overall coupling in Eq. (\ref{Eq:aeaN}) depends on both the axion mass and  the coupling  $g_{ae}$, we obtain a range for the  axion mass, see Fig. \ref{fig:avsgae}(a). 
  	 Having such a range of axion mass, it amusing to note that once the axion mass is determined one can determine the ratio $\beta=\frac{v_2}{v_1}$, see Fig.  Fig. \ref{fig:avsgae}(b).
  	 
\begin{figure}
  	\begin{center}
  			\rotatebox{90}{\hspace{-0.0cm} $R_0(m_a)\rightarrow$ counts per mol per year}
  			\includegraphics[width=0.5\textwidth]{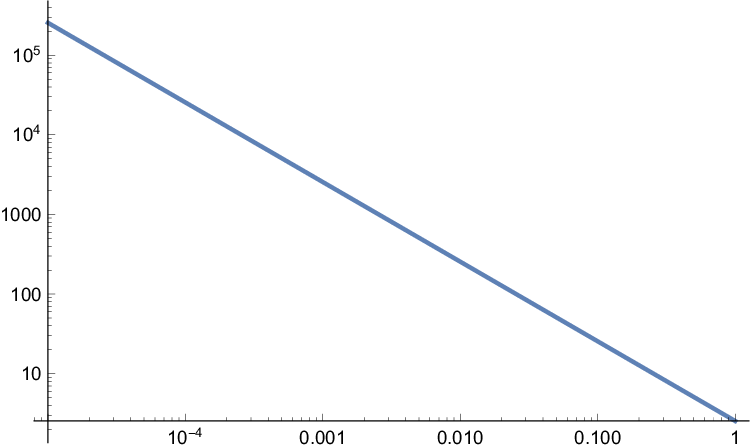}\\
  			{\hspace{-3.0cm}$m_a \rightarrow$ eV }
  		\caption{The scale of the rate as a function of the axion mass.}
  		\label{fig:R0(ma)}
  	\end{center}
  \end{figure}

  	 Anyway from Eq. (\ref{Eq:aeaN}), using  e.g.   $g_{ae}=\frac{1}{6}$,   one finds $f_a=5\times 10^{10}$ GeV. Using this information one finds the scale of the cross to be $\sigma_0=2\times 10^{-47}\mbox{cm}^2$. Furthermore for $N=6.0 \times 10^{23}$ atoms in one mol and the axion flux in our vicinity being  $\Phi_0=\tilde{\Phi}_0/m_a$ with $\tilde{\Phi}_0=2.0\times 10^{23}\frac{\mbox{1 eV}}{\mbox{y (cm)}^2}$, one   finds
  	 one finds
  $R_0(m_a)=\left (N \Phi_0 \sigma_0\right )= 2.55 \frac{{\mbox{1eV}}}{m_a}$ per mol-y. Thus  the main axion mass dependence   of the event is as  shown in  Fig, \ref{fig:R0(ma)}.

   Additional axion mass dependence, which can exploited by experiment, is contained in  $F_{m_1,m_2}(X)$ through $X$, see Eq.  (\ref{Eq:X(Delta)}).
  %The latter increases with the axion mass, since, as we shall see below, the parameter $f_a$ is a decreasing function of the axion mass. 

\section{The axion absorption  widths}
\label{sec:axionwidth}
Since the axion is absorbed one expects the cross section to exhibit a resonance behavior. This is exhibited by considering the function $F^N_{m_1,m_2}(X)$  in the  variable $X$, which  depends on the  energy difference of the atomic levels, the axion mass end the velocity of the sun around the center of the galaxy. The energy difference depends,  of course,  on the magnetic quantum numbers $m_1$ and $m_2$  of the states involved.

\begin{figure}
  \begin{center}
%	\subfloat[]	{
\rotatebox{90}{\hspace{-0.0cm} $F^N_{m_1,m_2}(X)\rightarrow$}
\includegraphics[width=0.5\textwidth]{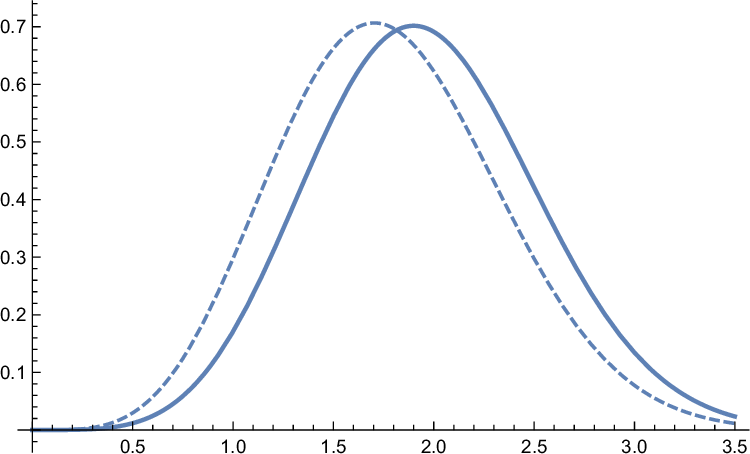}\\
{\hspace{-3.0cm}$X \rightarrow$  }
\\
%	\subfloat[]
%	{
\rotatebox{90}{\hspace{-0.0cm} $F(X)\rightarrow$}
\includegraphics[width=0.5\textwidth]{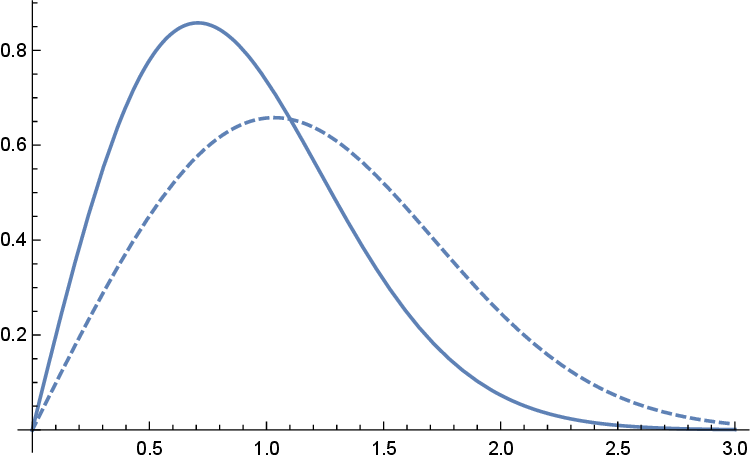}\\
{\hspace{-3.0cm}$X \rightarrow$  }\\
%}\\
%{\hspace{-2.0cm}$X \rightarrow$  }
%{\hspace{2.0cm}$X \rightarrow$  }
 \caption{Top panel: The normalized distribution $F^N_{m_1,m_2}(X)$ as a function of $X$, $X=\frac{c}{\upsilon_0}\left( \sqrt{2\left(\frac{E_f-E_i}{m_a c^2}-1\right )}\right)$ with $\upsilon_0$ the velocity of the sun around the center of the galaxy in natural units, $0.7 \times 10^{-3}$.
 %	220 km/s%, is exhibited. 
 	%The thick solid line corresponds to a Maxwell-Boltzmann distribution with respect to the galactic center.
 	 The solid line holds for  for $m_1=m_2$, while the dashed line for $m_2=m_1\pm 1$. The widths are  the same $\Gamma=1.35$  for both cases. The corresponding values of $\langle X \rangle $ are  1.9 and 1.7 for the solid and dashed curve respectively.
Bottom panel: For comparison the normalized distribution $F(X)$ $X=\frac{c}{\upsilon_0}\left( \sqrt{2\left(\frac{\omega}{m_a c^2}-1\right )}\right) $, with $\omega$ the photon energy, in the case of the standard axion to photon conversion is presented, obtained with the same halo parameters as in the top panel, in the galactic frame (solid curve) and local frame.
(dashed curve)}
 \label{fig:velfac}
 \end{center}
  \end{figure}

% Furthermore there exists an additional   dependence of $F(X)$ on the magnetic quantum numbers involved, which is mild.
 The overall behavior of the functions $F^N_{m_1,m_2}(X)$  is exhibited in Fig. \ref{fig:velfac}.
In fact we find that the  characteristics of the resonance are:
\beq
\left \{ \begin{array}{ccc} \Gamma=1.35, & \langle X \rangle=1.9,& \mbox{local frame, } m_1=m_2\\ \Gamma=1.35, & \langle X \rangle=1.7,& \mbox{local frame, } m_1\ne m_2\\ \end{array} \right . ,
\label{Eq:widthExpr}
\eeq
where  $\langle X \rangle$ is the location of the maximum 

% It is clear that, since the axion disappears, the excitation energy must be larger than the axion mass. It is, however, bounded by the maximum energy of the axion
%\beq
%m_a<E_f-E_i\le m_a(1+\frac{1}{2}v^2_{max})=m_a(1+\frac{1}{2}v^2_{esc})=m_a(1+\frac{1}{2}v_0^2 y^2_{esc})= m_a(1+\frac{1}{2}v_0^2 2.84^2)\approx m_a(1+2\times %10^{-6})
%\eeq

%This dependence of the widths on the magnetic quantum numbers of the states involved is a characteristic feature of the process and it can be exploited by experiments, if they reach adequate sensitivity in the measurement of the parameters of the width. Unfortunately the difference  in the expected widths, Eq. (\ref{Eq:widthExpr}), is small, about $1\%$, but the  location of the resonances can differ by more  than 10$\%$ (see Fig. \ref{fig:velfac}). In the latter case one needs to correct for the  dependence of $X$ on $\Delta m $, since  $\Delta m $ is not the same for both types of transitions.
	
\begin{figure}
  \begin{center}
\rotatebox{90}{\hspace{-0.0cm} $ F\left (\frac{c}{\upsilon_0}\sqrt{ 2\left (\frac{ E_f-E_i}{m_{a}c^2}-1\right )}\right )\rightarrow$}
\includegraphics[width=0.8\textwidth,height=0.3\textwidth]{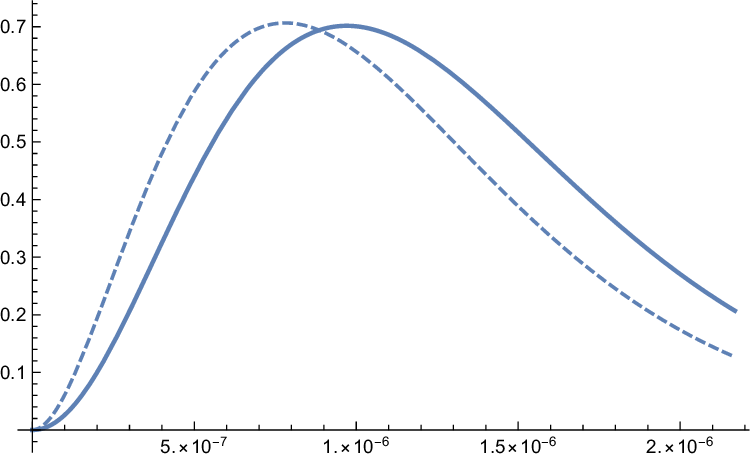}\\
{\hspace{-3.0cm}$ \left (\frac{ E_f-E_i}{m_{a}}-1\right )\rightarrow$  }
 \caption{The cross section exhibits resonance behavior. Shown is $F(X)$ as a function of $ \left (\frac{ E_f-E_i)}{m_{a}c^2}-1\right )$. The solid line corresponds to $m_1=m_2$, while the dashed line corresponds to $m_1\ne m_2$, both in the local frame. 
 \label{fig:axionefac}}
 \end{center}
  \end{figure}

	We have seen that we have  resonance behavior in the  variable $X$. At the location of the maximum from the relation   $\langle X \rangle =\frac{c}{\upsilon_0}\left( \sqrt{2\left(\frac{E_f-E_i}{m_a c^2}-1\right )}\right)$ we find that

	\beq 
	m_a=(E_f-E_i)\left (1-\chi \right ), \chi= 0.27 \times 10^{-6} \langle X \rangle^2 \mbox{ i.e } \begin{array}{cc}\chi= 0.97 \times 10^{-6},& m_1=m_2\\ \chi= 0.78 \times 10^{-6},& m_1\ne m_2\\
   \end{array}
   \label{Eq:resCond}
    \eeq
    For all practical purposes the axion mass is equal to the excitation energy. Furthermore
    	\beq 
    \frac{E_f-E_i}{m_a}=\left (1+\chi \right ), \chi= 0. 27 \times 10^{-6} \langle X \rangle ^2\mbox{ i.e } \begin{array}{cc}\chi= 0.97 \times 10^{-6},& m_1=m_2\\ \chi= 0.88 \times 10^{-6},& m_1\ne m_2\\	
  \end{array}
   \eeq
	Similarly we can ind the width in the energy space  for both types of transitions. Thus
	 $$ X_{1}=1.067 \leftrightarrow \left( \frac{E_f-E_i}{m_a}\right )_1=\left (1+ 0.306\times 10^{-6}\right ),\, X_{2}=2.417 \leftrightarrow \left( \frac{E_f-E_i}{m_a}\right )_2=\left (1+ 1.566\times 10^{-6}\right )\Rightarrow$$
	\beq
\Gamma_E=\left( \frac{E_f-E_i}{m_a}\right )_2-\left( \frac{E_f-E_i}{m_a}\right )_1=\left (\left (1+ 1.566\times 10^{-6}\right )-\left (1+ 0.306\times 10^{-6}\right )\right) = 1.26 \times 10^{-6}
%	\frac{E_f-E_i}{m_a}=\left (1+\chi' \right ), \mbox{ with } \begin{array}{cc}\chi'= 0.40 \times 10^{-6},& m_1=m_2\\ \chi'= 1.2 \times 10^{-6},& m_1\ne m_2\\
%	\end{array}
    \label{Eq.Ewidth}
	\eeq
	At $X=0$ we find  that at   $\frac{E_f-E_i}{m_a}=1$ the distribution vanishes. On the other hand  at $X=2.84$ i.e. for $\frac{E_f-E_i}{m_a}=1+2 \times 10^{-6}$, the distribution almost vanishes\footnote{There is no need to go to values of $X>y_{esc}=2.84$, since  velocities above the escape velocity, $\upsilon >\upsilon_{esc}=y_{esc}\upsilon_0$, in the Maxwell-Boltzmann distribution have been excluded.}. In other words the distribution vanishes at $X=0$ and at $X=2.84$ after having gone very rapidly through  $\langle X \rangle$ with a width as given by Eq. (\ref{Eq.Ewidth}). 	Anyway the above picture emerges more clearly  by plotting   the function $F(X)$ as a function of the the energy $(E_f-E_i)/m_ac^2$, see Fig.	\ref{fig:axionefac}.

	% Regarding the energy parameters we find
	%\beq
	%\Gamma_{E}=m_ac^2\left (1+\frac{1}{2} \left (\Gamma\frac{\upsilon^2_0}{c^2}\right )\right ),\,\langle E\rangle=m_ac^2\left (1+\frac{1}{2} \left (\langle X\rangle\frac{\upsilon^2_0}{c^2}\right )\right )
	%\eeq
	%Both are very close to the axion mass, since $(\upsilon_0/c)^2\approx0.5 \times 10^{-6}$\\
 We have seen that the axion mass is very close to the excitation energy. Since the resonance is so narrow, however, special care is required not to miss it. Some experimental arrangement to to facilitate the  observation of such a narrow resonance from the expected atomic spectra will be considered below, see section  \ref{sec:Rates}. Furthermore  oo this end the experience gained  with the axion to photon conversion experiments involving resonance cavities, such as ADMX and ADMX-HF collaborations 
	\cite{ExpSetUp11b},\cite{ADMX10},\cite{Stern14,MultIBSExp} and CAPP  \cite{CAPP}, \cite{ExpSetUp11a}, \cite{Semer20a},\cite{Semer20b},\cite{SemerArch19}, may be very helpful. The observation of the resonance is very important, among other things, to discriminate against background. It is very unlikely that background events will simulate a similar resonance pattern with that  that obtained here, reflecting not only the Maxwell-Boltzmann distribution, but the momentum dependence of the axion electron system as well. Furthermore, given enough counts, one can exploit,  if it becomes necessary, the extra signature,  provided by the fact that 
	the resonance width exhibits time dependence, i.e.  an annual modulation due to the motion of Earth, see \cite{VerSem16} and the Appendix, section \ref{sec:Appendix}.

% Before proceeding further  we should present the function $F^N_{m_1,m_2}(X)$ as a function of $\Delta=E_f-E_i$ by combining Eqs (\ref{Eq:F(X)}) and (\ref{Eq:X(Delta)}). 
Before completing this section we should mention  that resonance   plots, correlated with the event rate,   will be specialized in section  	\ref{sec:Rates} for the various atomic targets
considered in this work.

%	\subsection{Modulated widths due to the annual motion of the Earth}

\section{Atomic physics considerations}
\label{sec:AtomicPhys}
We will consider some atomic targets which possess two features. The ground state is composed of a multiplet of 
the form $n^{2 S+1}L_{J_1},S\ne 0$ while the excited state is of the form $n^{2 S+1}L_{J_2}$ with $|J_1-1|\le J_2\le J+1$, so that it can be reached by spin excitations.\\
The following types of excitations
 % |J_1,M_1\rangle\rightarrow|J_2,M2\rangle$ 
 are in principle possible:
\begin{itemize}
\item $|J_1,-J_1\rangle\rightarrow|J_1,-J_1+1\rangle$, indicated as type A
\item $|J_1,-J_1\rangle\rightarrow|J_2,-J_1-1\rangle$, indicated as type B
\item $|J_1,-J_1\rangle\rightarrow|J_2,-J_1\rangle$, indicated as type C
\item $|J_1,-J_1\rangle\rightarrow|J_2,-J_1+1\rangle$, indicated as type D
\end{itemize}
As we will see below some of these types may not be allowed by the angular momentum selection rules. Thus, e.g., for states $s_{1/2}$ single particle states only the $A$ type is possible. For many electron configurations see subsection  \ref{sec:ManyelConf}.
\subsection{Single particle spin orbit partners}
The ground state is a single particle state is of the form  $n\ell j_1,\,j_1=|\ell-1/2|$ while the state $n\ell j_2,j_2=\ell+1/2$ is empty. Such examples can be found in the atomic data.\\
Since this is an one body transition, $J_1=j_1,\,J_2=j_2$  the relevant matrix element takes the form:
\beq
C_{\ell,j_1,m_1,\,j_2,m_2}=\langle j_1\,m_1,1\,m_2-m_1|j_2\,m_2\rangle\sqrt{(2 j_1+1)3}\sqrt{2 \ell+1}\sqrt{6}\left \{ \begin{array}{ccc}\ell&\frac{1}{2}&j_1\\\ell&\frac{1}{2}&j_2 \\0&1&1 \end{array}\right \} (-1)^{m_1-m_2}
\eeq
i.e. it simply expressed in terms of a Glebsch-Gordan coefficient and the nine- j symbol. We are interested in the case $j_1,m_1=j_1,-J_1$. The relevant coefficients are tabulated in table \ref{tab:tab1}.

i) First we will consider a target with the ground state being a single  $p_{1/2} $   orbital, while the $p_{3/2} $ is empty. Let us suppose that the spin orbit splitting is $\epsilon_p$. In the presence of a magnetic field the m-degeneracy is removed and the ground state is  in the state $|j_1,m_1\rangle=|1/2,-1/2\rangle$. Then we have the following spin induced transitions:
$$|1/2,-1/2\rangle \rightarrow |1/2,1/2\rangle,\,|1/2,-1/2\rangle \rightarrow |3/2,-3/2\rangle, |1/2,-1/2\rangle\rangle \rightarrow |3/2,-1/2\rangle,|1/2,-1/2\rangle \rightarrow |3/2,1/2\rangle$$ 
indicated as as above, i.e. A,B,C and D respectively. 
% To leading order  the spin $g_s$  factors  are $g_s=(2/3,4/3)$ for $p_{1/2} $ and $p_{3/2} $ respectively. 
 Thus the   transition energies  are
\beq
m=\left\{2 \delta ,\epsilon
-\delta ,\frac{\delta
}{3}+\epsilon ,\frac{5
	\delta }{3}+\epsilon
\right\},\, \epsilon=\epsilon_p
\label{Eq:mForporb}
\eeq
where we have included both the spin and orbital magnetic moments with  $\delta=B \mu_B  $ with $\mu_B$ the Bohr magneton and   $B$  the magnetic field. For a  field of 1T we find $\delta=5.788\times 10^{-5}$ eV, i.e.
\beq
\delta=5.788\times 10^{-5}\frac{B}{1\mbox{T}} \mbox{eV}
\eeq
\\
A good candidate for such a transition is $_{13}$Al, involving the orbitals $3p_{1/2} $ and $3p_{3/2}$. From existing tables (https://www.nist.gov/pml/atomic-spectra-database)
we find $\epsilon _p=0.0139$ eV. The  spin induced matrix elements are as follows 
$$
C=\{2/9, 4/3, 8/9, 4/9 \} \mbox{ for the } A,\,B,\,C,\,D\mbox{ respectively}
$$

ii) Next we will consider a target with the ground state  containing a single  $d_{3/2} $   orbital, while the $d_{5/2} $ is empty. Let us suppose that the spin orbit splitting is $\epsilon_d$. In the presence of a magnetic field the m-degeneracy is removed and the ground state is  in the state $|j_1,m_1\rangle=|3/2,-3/2\rangle$. 
%To leading order  the $g_s$ values are $g_s=(4/5,6/5)$ for $d_{3/2} $ and $d_{5/2} $ 
Then we have the following spin induced transitions:
$$|3/2,-3/2\rangle \rightarrow |3/2,1/2\rangle,\,|3/2,-3/2\rangle \rightarrow |5/2,-5/2\rangle, |3/2,-3/2\rangle \rightarrow |5/2,-3/2\rangle,|3/2,-3/2\rangle \rightarrow |5/2,-1/2\rangle$$ 
indicated again  as A,B,C and D respectively. Thus the excitation energies are:
\beq
m= 
\left\{\frac{8 \delta
}{5},\epsilon -\frac{2
	\delta }{5},\frac{6 \delta
}{5}+\epsilon ,\frac{16
	\delta }{5}+\epsilon
\right\},\, \epsilon=\epsilon_d
\label{Eq:mFordorb}
\eeq
where we have included both spin and orbital magnetic moment\\ 
Our best candidate found in the above reference is the target  $_{21}$Sc involving the  $3d_{3/2} \rightarrow 3d_{5/2}$ transitions  with $\epsilon _d=0.021$ eV. 
% $$\mbox{ https://www.nist.gov/pml/atomic-spectra-database}$$.
Other  candidates  can also be found in the same reference, e.g.:
 Z=39 (Y I, $4d_{3/2}d_{5/2}$, 0.066 eV ) and Z=71 (Lu I, $5d_{3/2},d_{5/2}$, ~0.25 eV)
where I indicates that it is a neutral atom. \\
%A good candidate for such a transition is $_{13}$Al. Wi find $\epsilon _p=0.65$ which in good agrrement with existing tables (https://www.nist.gov/pml/atomic-spectra-database)
We thus  we can use Eq. (\ref{Eq:mFordorb}) with the appropriate value of $\epsilon_d$ and  
  the spin induced $|\mbox{ME}|^2$
$$C=\{ 4/25, 8/5, 16/25, 4/25 \}$$
%where again $C$ are the corresponding spin matrix elements.

iii) $s_{1/2}$ states. Such states  exist in many atomic  targets. In all such cases
$$ m= 2\delta,\, C=2.$$
We note the relatively  large spin matrix for single particle excitations.

 Note that in the case of $s_{1/2}$ and the A type transitions the lowest value of the   WIMP mass required for the process to take place is very small, since the spin orbit splitting does not appear. If such configuration exists in the ground state of the atom considered, the obtained results are independent of the atom.\\
The transition energy is also small for the other type of transitions, if the spin orbit splitting is small as, e.g., in the case for
all 3d-transitions considered here.
% On the other hand  in the case of 2p-levels  for the B,C,D transitions, a value of the mass  $\ge 0.0069$ eV is required, due to the fact that the spin orbit splitting is a bit higher (0.0069 eV). 
%The lower transition energies involved lead to some enhancement of the rates for the 3-d transitions as compared to those in the case of the 2p ones.

\subsection{More than one electron configurations}
\label{sec:ManyelConf}
%The needed matrix element is now given by
%\beq
%C_{L,J_1,m_1,\,J_2,m_2}=\langle J_1\,m_1,1\,m_2-m_1|J_2\,m_2\rangle\sqrt{(2 J_1+1)3}\sqrt{2 L+1}\langle S||\sbf||S\rangle\left \{ %\begin{array}{ccc}L&\frac{1}{2}&j_1\\L&\frac{1}{2}&j_2 \\0&1&1 \end{array}\right \} (-1)^{m_1-m_2}
%\eeq
%where $\langle S||\sbf|S \rangle$, which depends on the number of electrons in the configuration and $L$ the orbital angular momentum of the multiplet.

\subsubsection{Two electron configurations}
\label{sec:TwoEl}
The simplest possible case is two electron configurations. 
Now the needed  states are spin symmetric. Antisymmetry of the wave functions requires the space part to be antisymmetric, i.e. a wave function of the form $$\psi=\phi^2_{n\ell}(r)\left [L=\mbox{odd,}S=1\right]J=L-1,L,L+1$$
i.e.   spin triplet and  $L$ odd states. \\
Of special interest are the cases involving the functions:
$$\psi=\phi^2_{n\ell}(r)^3P_J,\quad \phi^2_{n\ell}(r)^3F_J$$
Then the spin matrix element can be cast in  form:
\beq
C_{L,J_1,m_1,\,J_2,m_2}=\langle ^3L_{J_2m_2}|\sigma|^3L_{J_1m_1}\rangle=\frac{1}{\sqrt{2J_2+1}}\langle J_1m_1,1m_2-m_1|J_2m_2\rangle 
\langle^3L_{J_2}||\sigma||\,^3L_{J_1}\rangle,\,L=P,F
\eeq
Good candidates  are the folowing:\\
i) $L=1$.\\
In this case the gs  of the carbon atom is of  the form $2s^2 2p^2\, ^3 P_0$, while the excited state which can be populated by spin excitations is  $^3 P_1$  at  16.41671 cm$^{-1}$, about 0.002 eV. It may be useful to note that the silicon atom  (Si I)  has the same structure, except for the radial quantum number $n=3$ and the fact that $\epsilon=0.00956$ eV. The first does not affect the  calculations performed here, while the second can be selected on the basis of the axion mass is being searched. That being said, the experimenters can choose whichever is more appropriate for them.\\
So since in both cases the  initial state is not degenerate, the only  excitations are $0\rightarrow m$ caused by the spin $\sigma_m$ appear.
%For our purposes this can be treated as a two particle system $2s^2 2p^2\, ^3 P_0$ due to particle hole symmetry. 
Furthermore one need consider of the  splitting of the final multiplet $2s^2 2p^2\, ^3 P_1$,  which is given by 
$$\left\{ -\frac{3	\delta }{2},0,\frac{3\delta }{2}\right\}\mbox { for }m=-1,\,0,\,1  \mbox{ respectively}.$$ Thus in the case of carbon
\beq
m=  \left\{ -\frac{3\delta }{2}+\epsilon,0,\frac{3\delta }{2}+\epsilon \right\}  ,\, \epsilon=0.002 \mbox{ eV} 
\label{Eq:carbonm}
\eeq

% The evaluation of the matrix element is a bit more complicated, but it can be calculated using standard techniques. Of special interest to us is the Fe atom. In this case the ground state is a six electron configuration 
 ii) $L=3$\\
 A good such candidate is  the Ti atom.  In this case
 the gs is of the form $4s^2 3d^2 \,^3 F_2$. The excited state that can be reached is $^3 F_3$ at 170.134, cm$^{-1}$=0.02 eV. Another good candidate is  the neutral Zirconium  (Zr I) atom. This  has the same structure, except for the radial quantum being $n=5$, which is irrelevant for our calculations, and the fact that $\epsilon=0.0707$ eV. The latter affects the axion mass to be extracted by the experimenters. So the choice of the target can be selected on the basis of the same criteria as above.
 Thus for both cases
 \beq
m= \left\{\frac{28 \delta
 }{9},\frac{121 \delta
 }{72}+\epsilon ,\frac{115
 	\delta }{36}+\epsilon
 ,\frac{113 \delta
 }{24}+\epsilon \right\},\,
% \{1.914 \delta , 2.631 \delta
% +\epsilon ,2.392 \delta
% +\epsilon ,2.153 \delta
% +\epsilon \},\, 
 %\epsilon=0.02 \mbox{ eV}
 \label{Eq:Tim}
 \eeq
 including both the spin and the orbital magnetic moment  with $\epsilon$ the two values just mentioned.
 
 The relevant spin matrix element is given in table \ref{tab:tab2}.
 \subsubsection{More than two electron configurations}
 \label{sec:ManyPartCon}
 i) The oxygen atom.\\
 In this case the gs is of the form $2s^2 2p^4\, ^3 P_2$, while the excited state which can be populated by spin excitations is  $^3 P_1$  at  158.265 cm$^{-1}$, about 0.0196 eV.  This target is appropriate at the low temperatures we consider, since it is no longer a gas. Otherwise one might consider the atom of Sulfur (S I) which has the same configuration but with n=3 instead of n=2. Returning to the oxygen atom,
 %So the initial state is not degenerate, so the only the excitations are $0\rightarrow m$ caused by the spin $\sigma_m$ appear. 
 %For our purposes this can be treated as a two particle system $2s^2 2p^2\, ^3 P_0$ due to particle hole symmetry. 
  we will  consider of the  splitting of the ground state  multiplet $2s^2 2p^4\, ^3 P_2$,  as well as the final state multiplet $2s^2 2p^4\, ^3 P_1$
 %$$\left\{ -\frac{3	\delta }{2},0,\frac{3\delta }{2}\right\}\mbox { for }m=-1,\,0,\,1  \mbox{ respectively}.$$ Thus
 %\beq
 %m=  \left\{ -\frac{3\delta }{2}+\epsilon,0,\frac{3\delta }{2}+\epsilon \right\}  ,\, \epsilon=0.0196 \mbox{ eV} 
 %\label{Eq:Oxygen}
 %\eeq
 
% The evaluation of the matrix element is a bit more complicated, but it can be calculated using standard techniques. Of special interest to us is the Fe atom. In this case the ground state is a six electron configuration 	

By angular momentum selection rules only the $A$ and the $D$ terms are allowed. Thus one finds
\beq
m=\left\{\frac{5 \delta
}{3},0,0,\frac{5 \delta
}{3}+\epsilon \right\},
%\left\{\frac{3 \delta
%}{2},0,0,\frac{3 \delta
%}{2}+\epsilon \right\},
 \epsilon=0.0196\mbox{ eV}.
\eeq
including the contribution of both the spin and orbital magnetic moment.

 ii) The $_{26}$Fe atom.\\	 		 		 		 		 		 		 		 		 
 $3d^6 4s^2\,^5D_4$ while the excited state that can be populated by spin excitations is the first excited state $^5D_3$ at 415.933cm$^{-1}$=0.05 eV.
%The reduced matrix elements are given in table \ref{tab:tab2}. 
% The reduced matrix elements  and the full matrix element for the two m-combinations allowed in the case  $\langle ^5D_{J_2m_2}|\sigma|\,^5D_{J_1,-J1}\rangle^2$ are given in table \ref{tab:tab2}.\\
The calculation of the reduced spin and orbital angular momentum  matrix elements is a bit complicated, but it can be simplified by making use of the symmetries of the wave functions. The spin part is characterized by the SU(2) symmetry $[5,1]$ while antisymmetry requires the orbital part to be $[2,1^4]$ 
%This is equivalent to the  configurations $3d^4 4s^2\,^5D_4$ and $3d^4 4s^2\,^5D_3$ for the ground and the excited state respectively. Thus the is equivalent to a symmetric spin state indicated as [4] and a completely antisymemtric orbital state of the type $[1,1,1,1]L=2$ 
under SU(5). The relevant matrix elements can be evaluated using standard techniques, see e.g \cite{JDV17}, making use of table B.19. 
% Taking half of the relevant reduced spin matrix elements and $g_s=2$  one can evaluate the needed matrix elements of the  magnetic moment and the resulting splitting.
 Since  only the transitions  of the type A and D appear in this case, one finds
 \beq
 m=\left\{\frac{16 \delta
 }{5},0,0,\frac{16 \delta
 }{5}+\epsilon \right\},
 %m=\{12.0448 \delta.
 %,0.,0.,19.7142 \delta
 %+\epsilon \},
 % \left \{\frac{\delta}{2},0,0,(\epsilon+\delta)\frac{1}{2} \right \}
  \epsilon=0.05 \mbox{ eV}
  \label{Eq:ironm}
 \eeq
 including both the spin and the orbital magnetic moments.
 
 The needed spin induced transition matrix elements $\langle ^3P_{J_2m_2}|\sigma|\,^3P_{J_1,-J1}\rangle^2$ and $\langle ^3F_{J_2m_2}|\sigma|\,^3F_{J_1,-J1}\rangle^2$ and $\langle ^5D_{J_2m_2}|\sigma|\,^5D_{J_1,-J1}\rangle^2$  are shown  in \ref{tab:tab2}

%The reduced matrix elements are given in table \ref{tab:tab2}. the full matrix element $\langle ^3P_{J_2m_2}|\sigma|\,^3P_{J_1,-J1}\rangle^2$ is also shown  in \ref{tab:tab2}
%For orientation purposes we will take an average value for $\langle X\rangle =1.8$ for $m_1=m_2$ type excitations and 1.6 for $m_1\ne m_2$. 
%The obtained results for $\ell=0$ and $\ell=1$ are exhibited in table \ref{tab:results}.
\begin{table}
\caption{The coefficients $\left (C_{j_1,m_1,j_2,m_2,\ell}\right)^2 $ connecting via the spin operator a given initial state
 $|i\rangle=|n\ell,j_1,-j_1\rangle$ with all possible states $|f \rangle=|n\ell,j_2,m_2\rangle $, for $\ell=0,\,1,\,2,\,3$.}
\label{tab:tab1}
$$
\left (
\begin{array}{ccccc|c}
&|i\rangle&&|f\rangle&&\\
\hline
\ell&j_1&m_1&j_2&m_2&C^2_{j_1,m_1,j_2,m_2,\ell}\\
\hline
0 & \frac{1}{2} & -\frac{1}{2} & \frac{1}{2} & \frac{1}{2} &
2 \\
 1 & \frac{1}{2} & -\frac{1}{2} & \frac{1}{2} & \frac{1}{2} &
   \frac{2}{9} \\
 1 & \frac{1}{2} & -\frac{1}{2} & \frac{3}{2} & -\frac{3}{2} &
   \frac{4}{3} \\
 1 & \frac{1}{2} & -\frac{1}{2} & \frac{3}{2} & -\frac{1}{2} &
   \frac{8}{9} \\
 1 & \frac{1}{2} & -\frac{1}{2} & \frac{3}{2} & \frac{1}{2} &
   \frac{4}{9} \\
 2 & \frac{3}{2} & -\frac{3}{2} & \frac{3}{2} & \frac{1}{2} &
   \frac{6}{25} \\
 2 & \frac{3}{2} & -\frac{3}{2} & \frac{5}{2} & -\frac{5}{2} &
   \frac{8}{5} \\
 2 & \frac{3}{2} & -\frac{3}{2} & \frac{5}{2} & -\frac{3}{2} & \frac{16}{25}
   \\
 2 & \frac{3}{2} & -\frac{3}{2} & \frac{5}{2} & -\frac{1}{2} &
   \frac{4}{25} \\
 3 & \frac{5}{2} & -\frac{5}{2} & \frac{5}{2} &- \frac{3}{2} &
   \frac{10}{49} \\
 3 & \frac{5}{2} & -\frac{5}{2} & \frac{5}{2} &- \frac{7}{2} & \frac{12}{7}\\
  3 & \frac{5}{2} & -\frac{5}{2} & \frac{5}{2} &- \frac{5}{2} &\frac{24}{49}\\
   3 & \frac{5}{2} & -\frac{5}{2} & \frac{5}{2} &- \frac{3}{2} &\frac{4}{49}\\
   \\
\end{array}
\right)
$$
\end{table}

\begin{table}
	\caption{The coefficients $ \langle^3P_{J_2}||\sigma||\,^3P_{J_1}\rangle$,  $ \langle^3F_{J_2}||\sigma||\,^3F_{J_1}\rangle$ and $\langle^5D_{J_2}||\sigma||\,^5D_{J_1}\rangle$ as well as the corresponding total matrix elements $|ME|^2$ relevant to the present work. Note that the initial sub-state is of the form $|j_1,m_1\rangle =|J_1,-J_1\rangle$ }
	\label{tab:tab2}
	$$
	\begin{array}{cccccc|c}
	J_1&J_2&\langle^3P_{J_2}||\sigma||\,^3P_{J_1}\rangle&m_1&q&\frac{\langle J_1,m_11,1 q|J_2,m_1+q\rangle}{\sqrt{2 J_2+1}}&|ME|^2\\
	\hline
	0 & 1&2 \sqrt{2}&0 & \pm 1,0&\frac{1}{\sqrt{3}} & 8/3\\
	\hline
%	\hline
%	1 & 0&-2 \sqrt{6}&-1&1&\frac{1}{\sqrt{3}}&8\\
%	1 & 1&3 \sqrt{2}&-1&1&-\frac{1}{\sqrt{6}}&3\\
%	1 & 2& \sqrt{30}&-1&-1&\frac{1}{\sqrt{5}}&6\\
%	1 & 2& \sqrt{30}&-1&0&\frac{1}{\sqrt{10}}&3\\
%	1 & 2& \sqrt{30}&-1&1&\frac{1}{\sqrt{30}}&1\\
%	2 & 1& -\sqrt{30}&-2&1&-\frac{1}{\sqrt{5}}&6\\
%	2 & 2& -\sqrt{30}&-2&1&-\frac{1}{\sqrt{15}}&2\\
	\end{array}$$
	$$
		\begin{array}{cccccc|c}
	J_1&J_2&\langle^3F_{J_2}||\sigma||\,^3F_{J_1}\rangle&m_1&q&\frac{\langle J_1,m_11,1 q|J_2,m_1+q\rangle}{\sqrt{2 J_2+1}}&|ME|^2\\
	\hline
	2 & 2&-2 \sqrt{\frac{10}{3}}&-2 &  1&-\frac{1}{\sqrt{15}} & \frac{8}{9}\\
	2 & 3&-4\frac{\sqrt{5}}{\sqrt{3}} &-2&-1&\frac{1}{\sqrt{7}}&\frac{80}{21}\\
	2 & 3&-4\frac{\sqrt{5}}{\sqrt{3}} &-2&0&\frac{1}{\sqrt{21}}&\frac{80}{63}\\
	2 & 3&-4\frac{\sqrt{5}}{\sqrt{3}} &-2&1&\frac{1}{\sqrt{105}}&\frac{16}{63}\\
		\hline
%	\hline
%	3 & 2&-\frac{20}{\sqrt{3}} &-3&1&\frac{1}{\sqrt{7}}&\frac{400}{21}\\
%	3 & 3&\sqrt{\frac{35}{3}} &-3&1&\frac{1}{2\sqrt{7}}&\frac{5}{12}\\
%	3 & 4&3\sqrt{15} &-3&-1&\frac{1}{3}&15\\
%	3 & 4&3\sqrt{15} &-3&0&\frac{1}{6}&\frac{15}{4}\\
%	3 & 4&3\sqrt{15} &-3&1&\frac{1}{6\sqrt{7}}&\frac{15}{28}\\
%	4 & 3&-3\sqrt{15} &-4&1&\frac{1}{3}&15\\
%	4 & 4&15 &-4&1&-\frac{1}{3\sqrt{5}}&5\\
	\end{array}
	$$
	$$
			\begin{array}{cccccc|c}
	J_1&J_2&\langle^3P_{J_2}||\sigma||\,^3P_{J_1}\rangle&m_1&q&\frac{\langle J_1,m_11,1 q|J_2,m_1+q\rangle}{\sqrt{2 J_2+1}}&|ME|^2\\
	\hline
	2 & 2& {\sqrt{30}}&-2 &  1&-\frac{1}{\sqrt{15}} & 2\\
	2 & 1 &\sqrt{10}  &-2 & 1&\frac{1}{\sqrt{5}} &2\\
		\hline
	\hline
%	3 & 3&\sqrt{21} &-3 &  1&-\frac{1}{2\sqrt{7}} & \frac{3}{4}\\
	\end{array}
	$$
	$$
	\begin{array}{cccccc|c}
	J_1&J_2&\langle^5D_{J_2}||\sigma||\,^5D_{J_1}\rangle&m_1&q&\frac{\langle J_1,m_11,1 q|J_2,m_1+q\rangle}{\sqrt{2 J_2+1}}&|ME|^2\\
	\hline
	4 & 4& 6{\sqrt{5}}&-4 &  1&-\frac{1}{3\sqrt{5}} & 4\\
	4 & 3 &6  &-4 & 1&\frac{1}{3} &4\\
	\hline
	\hline
	%	3 & 3&\sqrt{21} &-3 &  1&-\frac{1}{2\sqrt{7}} & \frac{3}{4}\\
	\end{array}
	$$
\end{table}

\section{Low temperature requirements}
\label{ec:lowtemp}
As we have mentioned the  detection of very light axions, in the regime of a few $\mu$eV mass, crucially depends on the condition that the second level must be essentially free of electrons.
To achieve this condition  the target material should be brought at low temperatures. The critical temperature depends on the axion mass to be explored. The ratio of the probabiliies of finding an accidental electron in the second level relative to the probability of finding one in the first level id given by the Boltzmann distribution probability:
\beq
P_{i,f}=e^{-m_a/kT}
\eeq
Suppose that we demand this to be $10^{-x}$. Then we find that 
\beq
T\le \frac{0.434 m_a}{k x}
\eeq

\begin{figure}
	\begin{center}
		\rotatebox{90}{\hspace{-0.0cm} $T \rightarrow ^{0}$K}
		\includegraphics[width=0.8\textwidth,height=0.4\textwidth]{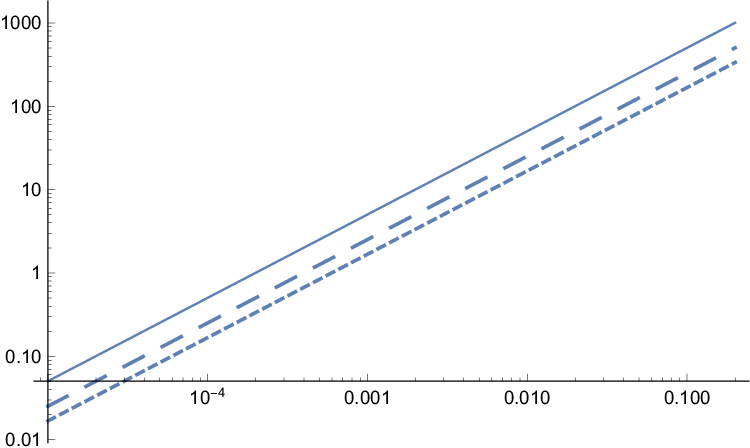}\\
		\hspace{-3.0cm}$m_a \rightarrow $eV  \\
		\caption{The temperature in degrees Kelvin to be achieved is the region below the above curves, so that the population of the excited state by thermal electrons can be neglected. the continues corve corresponds to relative probability of $10\%$, the long dash to $1\%$ and the sort dash to $0.1\%$}
		\label{fig:Tvsma}
	\end{center}
\end{figure}

The condition on the temperature is given in Fig. \ref{fig:Tvsma}. 
%	\section{Some experimental considerations}
We thus see that, for axion mass $m_a=0.4\times 10^{-4}$ eV associated with a magnetic field B=1T, $T\le 0.1 \,^0$K may be required. The situation may  improve, of course,  for larger magnetic fields. Anyway for axions  heavier than 0.1 eV, very low temperatures are not required. Accidental backgrounds causing the excitation may be rejected from the resonance behavior, since it is unlikely that they are going to have a velocity distribution similar to that of the axions in the local frame.

There remains, however, an additional problem. For for very light axions one has to develop detector materials, which at these low temperatures exhibit atomic structure.   Ordinary atoms do not suffice. The ions of the crystal still exhibit atomic structure involving the bound electrons, as, e.g., the CUORE detector of
Crystalline $^{130}$TeO$_2$ at low  temperatures. The
electronic states probably won't carry all the important quantum numbers as
their corresponding neutral atoms, but they should possess the configurations connected by the spin excitations considered here. So one may prefer to consider targets which  contain  appropriate impurity atoms in a host crystal, e.g chromium in sapphire. As a matter of fact it is very encouraging that already there exist  proposals involving  rare-earth
ions doped into solid-state crystalline materials \cite{Padova17} at low temperatures. In fact, if ion impurities are doped in crystals, one has to choose the target atoms in such a way that the corresponding ions can be isoelectronic to the atoms presented here and having the same configurations and terms. Since, as we shall see below,  the expected rates for light axions are quite large, small impurities of 1 to 1000 or even 1/10000 may be adequate or one may relax the condition on the relative probability  $P_{i,f}$.

It is, also, possible  that one may be able to employ at low temperatures some exotic materials used in quantum technologies (for a  review see \cite{NVReview13}) like 
nitrogen-vacancy (NV), i.e. materials characterized by spin $S=1$, which in a magnetic field  allow transitions between $m=0,\,m=1$ and $m=-1$. 
	\section{Some estimates on the expected rates}
	\label{sec:Rates}
	We have seen that the event rate is given by Eq. (\ref{Eq:rate}). Its scale, computed with    a cross section $\sigma_0=2.0\times 10^{-47}\mbox{cm}^2$, extracted from the Borexino data, is shown in  Fig, \ref{fig:R0(ma)}. 
We are now going to compute the total rates incorporating the spin induced matrix elements $\left (C_{j_1,m_1,j_2,m_2,\ell}\right)^2 $ and the effects of the resonance. Our results will be presented   in the form of suitable  plots.

  In the plots the resonance behavior will be  apparent, but the location of the resonance as well as its width depend on the atom considered through the parameters $m_i,\,i=A,B,C,D$, which are functions of  the spin orbit splitting $\epsilon$ as well as the energy $\delta$ due to the magnetic moment. These are given in section \ref{sec:AtomicPhys}. We should also indicate the  event rate on the resonance for each type of transition  $ A,B,C,D$, which will not be the same for all of them  due to the different axion  mass and the spin induced matrix elements. For compactness of presentation we will put all  this information for a given atomic target   in the same plot. The event rate will  be most economically presented for all transition types in the same figure as a function of $r=\frac{\Delta}{m_a}-1$, with  $\frac{\Delta}{m_a}$ covering the range of values allowed in the interval between $X=0$ and $X=2.84$, as  discussed in section \ref{sec:axionwidth},  in a fashion  analogous to that  of  Fig. \ref{fig:axionefac}. The picture in this case  will necessarily be more complicated, but we hope that,  with the information provided, it will be understood, after the discussion of section   \ref{sec:axionwidth}.\\
  
  % For a given type of transition the maximum occurs roughly at $\Delta=m_a$.

  Before we proceed further with the details of the rates of various atoms we should mention that the resonance condition given by Eq.  (\ref{Eq:resCond}) must be  satisfied. It can now be written as
  \beq 
  m_a=m_i\left (1-\chi \right ),  \begin{array}{cc}\chi= 0.97 \times 10^{-6},& i=C\\ \chi= 0.78 \times 10^{-6},& i=A,B,D\\
  \end{array}
  \label{Eq:ResCond2}
  \eeq
  %with $\Delta_=m_i$ 
  
 %  will make some qualitative estimates of the expected widths and locations of the resonances in the energy space. Using Eqs (\ref{Eq:widthExpr}) and (\ref{Eq:X(Delta)}) one finds
  %\begin{itemize}
  %	\item  $\langle E\rangle_A\approx 3\times 10^{-11}\frac{\mbox{B}}{1\mbox{T}}$, $\Gamma_A\approx 2\times 10^{-11}\frac{\mbox{B}}{1\mbox{T}}$
  %		\item  $\langle E\rangle_{B,D}\approx 8 \times 10^{-7}\frac{\epsilon}{\mbox{1eV}}$, $\Gamma_{B,D}\approx 5.5\times 10^{-7}\frac{\epsilon}{\mbox{1eV}}$
 % 		\item  $\langle E\rangle_{C}\approx 7 \times 10^{-7}\frac{\epsilon}{\mbox{1eV}}$, $\Gamma_A\approx 5\times 10^{-7}\frac{\epsilon}{\mbox{1eV}}$	
 % \end{itemize}
%We should keep in mind, however, that the rates are not normalized\\ 

\subsection{One electron configurations}

\begin{figure}
	\begin{center}
		\rotatebox{90}{\hspace{-0.0cm} $R \rightarrow $ per mol-year}
		\includegraphics[width=0.5\textwidth]{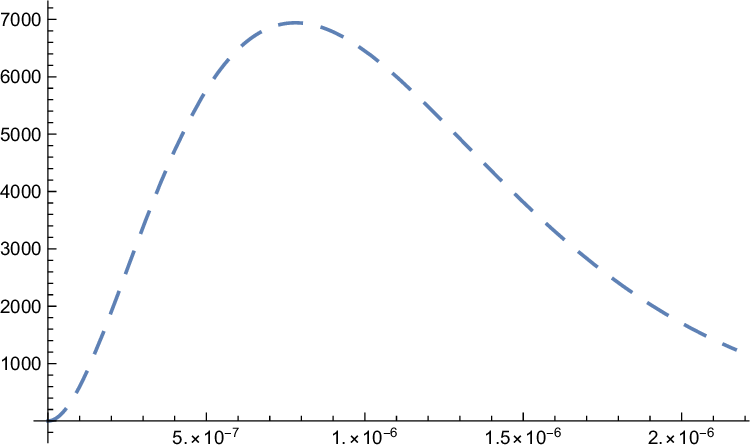}\\
		\hspace{-3.0cm}$r=\frac{\Delta}{m_a}-1 \rightarrow $  \\
		\caption{ The only possible transition is of $A$ type. Now the transition energy  is $\Delta=6 \times 10^{-5}\frac{B}{1\mbox{T}} (1+r)$ eV. The extracted neutrino mass is given by the value of $r_0$ at the location of the maximum, $m_a=M_A(1-r_0)$ , analogous to that of   Eq. (\ref{Eq:ResCond2}). The width is determined by $\Gamma=r_2-r_1$, where $r_2$ and $r_1$ the locations at half maximum.
			%The extracted width is $\Gamma\approx 10^{-11}$eV
		}
		\label{fig:sSP}
	\end{center}
\end{figure}

  We will consider the following cases:
  
%We will consider the following cases:
i) s-orbitals. \\For such atoms the obtained rate is shown in Fig. \ref{fig:sSP}.

ii) p-orbitals.

\begin{figure}
	\begin{center}
		\rotatebox{90}{\hspace{-0.0cm} $R \rightarrow $ per mol-year}
		\includegraphics[width=0.8\textwidth]{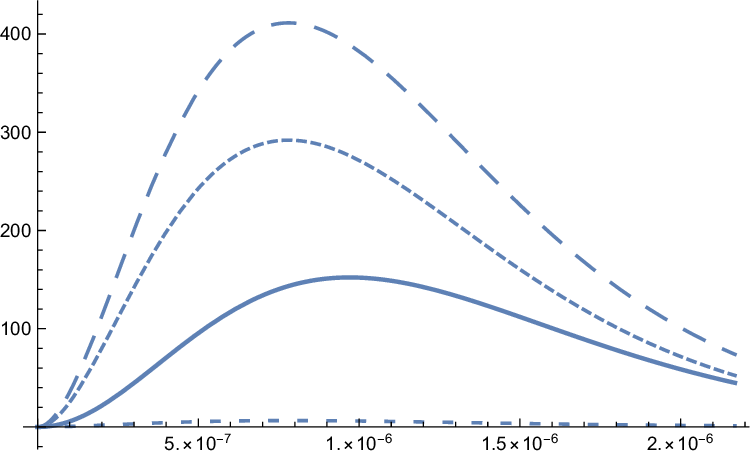}\\
		\hspace{-3.0cm}$r=\frac{\Delta_i}{m_a}-1 \rightarrow $  \\
		\caption{The rate as a function of $r=\frac{\Delta}{m_a}-1$, as a function of $r$, for transition types $A,\, B,\,C ,\, D$ indicated by long dash, short dash, solid line and intermediate dash respectively.  Note that to make the type $A$ fit in the picture we have suppressed it by the factor $C_g=1/5$, i.e. its actual value ia 5 times larger. In these plots the shape is  essentially determined the velocity distribution. the extraction of the axion mass  and the resonance width is determined  as in Fig.\ref{fig:sSP}, except that mow  
			$r_0\rightarrow r^i_0$, $r_1\rightarrow r^i_1$ and $r_2\rightarrow r^i_2$, $i=A,B,C,D$. The obtained results for the axion mass  and the extracted width may not agree exactly with those of Eq. Eq. (\ref{Eq:ResCond2}) and (\ref{Eq.Ewidth}) respectively due to the fact that the resonances here are not normalized.
			 %The actual transition energy  $\Delta$ is obtained from $r$ by $\Delta=m_i (1+r)$. $m_i$ is a short hand notation for $m_a(i), \, i=A,B,C,D $.
			 The parameters $m_i$  are as follows:
			$m_A=\frac{2 \delta
			}{3}$,  $m_B=\epsilon-\frac{5
				\delta }{3}$, $m_C=\epsilon
			-\frac{\delta }{3}$, $m_D=\delta
			+\epsilon$. In the case of 
			   $_{13}$Al considered here $\epsilon=0.0130$ eV. For any other   single particle p-orbitals only $\epsilon$ may be different.
		%	The resonance behavior is quite clear.
			% Its width is approximately given by Eq. (\ref{Eq.Ewidth}). Here, however, the resonance pattern is not normalized and, as a result, the difference of the locations at half maximum may differ,  depending on the parameters $m_i,\,i=A.B.C,D$. 
		%	i.e.  $\epsilon$ as well as  $\delta$ (proportional to the  magnetic field employed). Thus they become  different for the various types, especially for  A type  and the rest.
		 }
		\label{fig:pSP}
	\end{center}
\end{figure}

\begin{figure}
	\begin{center}
		\rotatebox{90}{\hspace{-0.0cm} $R \rightarrow $ per mol-year}
		\includegraphics[width=0.8\textwidth]{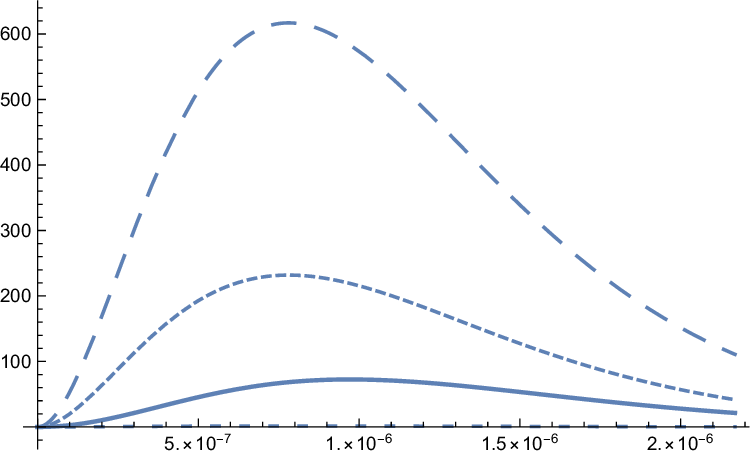}\\
		\hspace{-3.0cm}$r=\frac{\Delta}{m_a}-1 \rightarrow $  \\
		\caption{The same as in Fig. \ref{fig:pSP} in the case of the atom $_{21}$Sc. 
		%	The transition energy is obtained from $r$ by $\Delta=m_i (1+r),\, i=A,B,C,D$.
		 The parameters $m_i$  are as follows:		
			$m_A=\frac{8 \delta
			}{5}$, $m_B=\epsilon-\frac{9
				\delta }{5}$, $m_C=\epsilon
			-\frac{3 \delta
			}{5}$, $m_D=\epsilon+\frac{3
				\delta }{5}$, with $\epsilon=0.021$eV. 
		}
		\label{fig:dSc}
			\end{center}
		\end{figure}

\begin{figure}
	\begin{center}
		\rotatebox{90}{\hspace{-0.0cm} $R \rightarrow $ per mol-year}
		\includegraphics[width=0.8\textwidth]{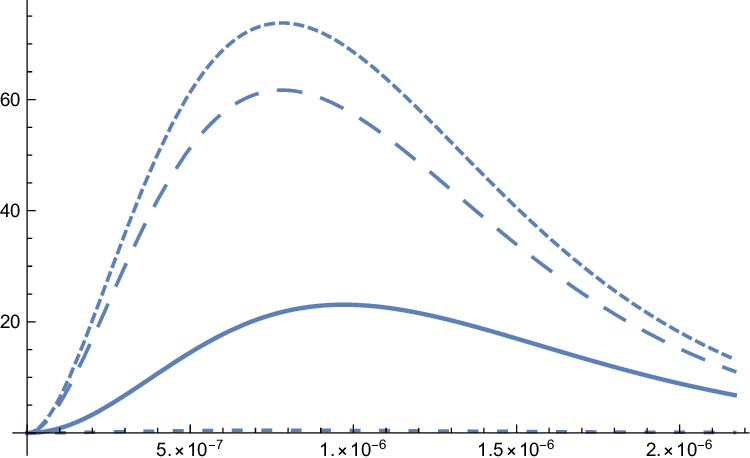}\\
		\hspace{-3.0cm}$r=\frac{\Delta}{m_a}-1 \rightarrow $  \\
		\caption{The same as in Fig. 	\ref{fig:dSc} in the case of the atom $_{39}$Y, which also involves d-orbitals, but in this case $\epsilon=0.066$eV . For transition type $ A$ the suppression factor $C_g=1/10$ was employed, i.e the corresponding rate must be multiplied by 10. 
		}
		\label{fig:dY39}
	\end{center}
\end{figure}

\begin{figure}
	\begin{center}
		\rotatebox{90}{\hspace{-0.0cm} $R \rightarrow $ per mol-year}
		\includegraphics[width=0.8\textwidth]{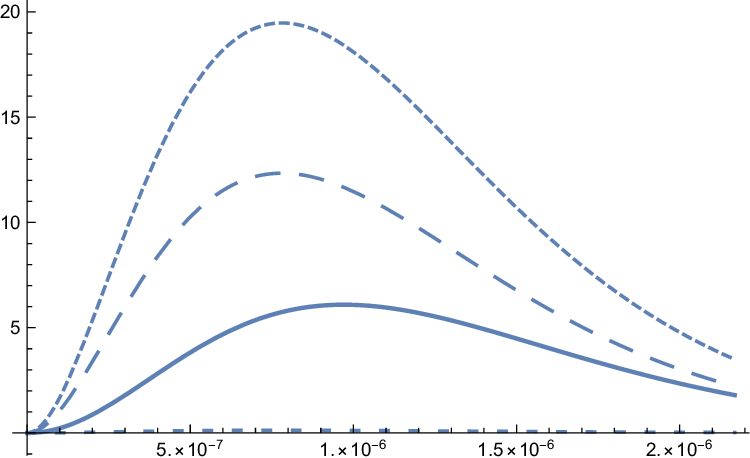}\\
		\hspace{-3.0cm}$r=\frac{\Delta}{m_a}-1 \rightarrow $  \\
		\caption{The same as in Fig. 	\ref{fig:dSc} in the case of the atom $_{71}$Lu, which also involves d-orbitals, but in this case $\epsilon=0.25$eV . Again  for transition type $ A$ the suppression factor $C_g=1/50$ was employed.
		}
		\label{fig:dLu71}
	\end{center}
\end{figure}

 The event rate is exhibited in Fig. \ref{fig:pSP}.

iii) d-orbitals.\\ The obtained event rates are shown  in Figs \ref{fig:dSc}-\ref{fig:dLu71}.


\subsection{Two electron configurations}

\begin{figure}
	\begin{center}
		\rotatebox{90}{\hspace{-0.0cm} $R \rightarrow $ per mol-year}
		\includegraphics[width=0.5\textwidth]{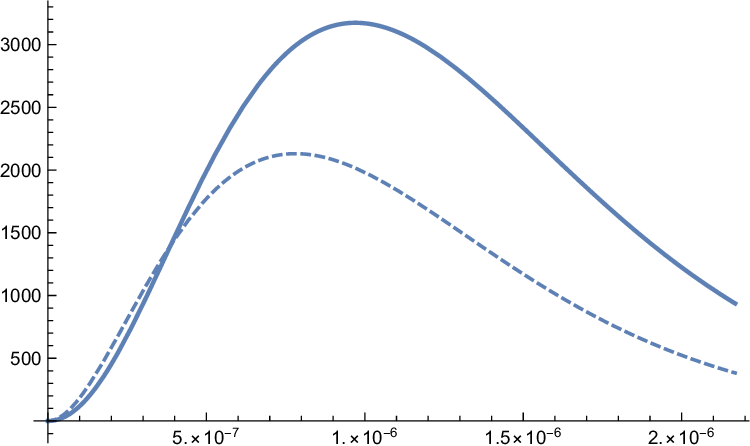}\\
		\hspace{-3.0cm}$r=\frac{\Delta_i}{m_a}-1 \rightarrow $  \\
		\caption{The same as in Fig. \ref{fig:pSP} but for the C atom, with $m_i$ as follows: 
			%Now the transition energy  is obtained from $r$ by $\Delta=m_i (1+r),\, i=B,C,D$ with the parameters $m_i$ given by
			$m_B=\epsilon-\frac{3}{2}\delta$, $m_C=\epsilon$, $m_D=\epsilon +\frac{3}{2}\delta $, $\epsilon=0.002$  eV .
			The patterns $B$ and $D$ coincide, while the type $ A$ transition  is not present.
			%	 For transition types $ B,\,C ,\, D$ the value $C_g=10^{-4}$ was used.
		}
		\label{fig:C2P}
	\end{center}
\end{figure}

In this case we will consider  the two systems discussed in section \ref{sec:TwoEl}, namely carbon and  $_{22}$Ti.\\
In the first case the transition is $^3P_0\rightarrow ^3P_1$, 
%The simplest such system is $_{6}$C.
%, see section \ref{sec:TwoEl}.
%The first  case the transition is $^3P_0\rightarrow ^3P_1$ 
 i.e,  since the initial  initial state is a $J=0$, the $A$ type transition is not available. The obtained rates are exhibited in Fig. \ref{fig:C2P}. It may be useful to note that the Si atom has the same structure, except for the radial quantum, which is irrelevant here, and the fact that $\epsilon=0.00956$ eV. Otherwise the situation is the same as in Fig. \ref{fig:C2P}

\begin{figure}
	\begin{center}
		\rotatebox{90}{\hspace{-0.0cm} $R \rightarrow $ per mol-year}
		\includegraphics[width=0.5\textwidth]{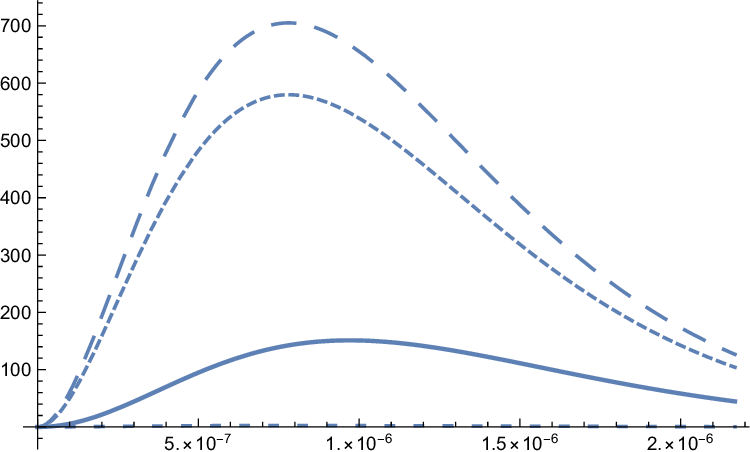}\\
		\hspace{-3.0cm}$r=\frac{\Delta}{m_a}-1 \rightarrow $  \\
		\caption{The same as in Fig. \ref{fig:pSP} but for the Ti atom. For transition type $ A$ the suppression factor $C_g=1/5$ was employed, i.e the corresponding rate must be multiplied by 5. Now  $m_i$ given by	
		$m_A=\frac{28 \delta
		}{9}$, $m_B=\frac{121 \delta
		}{72}+\epsilon $, $m_C=\frac{115
			\delta }{36}+\epsilon$
		, $m_D=\frac{113 \delta
		}{24}+\epsilon$,
		%	$ m_A=1.914 \delta$, 
		%	 $m_B=2.631 \delta
		%	+\epsilon$, $m_C=2.392 \delta
		%	+\epsilon$, $m_D=2.153
		%		\delta +\epsilon$, 
		 $\epsilon=0.02$  eV.  The 	The type $ D$ transition  is not visible.
			%For transition types $ B,\,C ,\, D$ the value $C_g=10^{-4}$ was used.
		}
		\label{fig:Ti22}
	\end{center}
\end{figure}

 The second target  $_{22}$Ti, the  transition is $^3F_2\rightarrow ^3F_3$,  with  the splitting of the spin orbit partners being relatively  small. The obtained results  are exhibited in Fig. \ref{fig:Ti22}. This use of such target, however, may suffer from the fact that  $_{22}$Ti normally   exists in metallic form. It may be useful to note that the neutral Zr (Zr I) atom has the same structure, except for the radial quantum number, which is irrelevant here, and the fact that $\epsilon=0.0707$ eV. Otherwise the situation is the same as in Fig. \ref{fig:Ti22}.

\subsection{Many electron configurations}
In this case we will consider the two targets described in section \ref{sec:ManyPartCon}.\\
i)  Transitions  of the type  $^3P_2\rightarrow ^3P_1$.\\
The simplest such system is  the oxygen  with $n=2$. As it has already mentioned, one may also  consider sulfur (S I), which has   the same configuration but different $n=3$.

%, see section \ref{sec:TwoEl}.

\begin{figure}
	\begin{center}
		\rotatebox{90}{\hspace{-0.0cm} $R \rightarrow $ per mol-year}
		\includegraphics[width=0.5\textwidth]{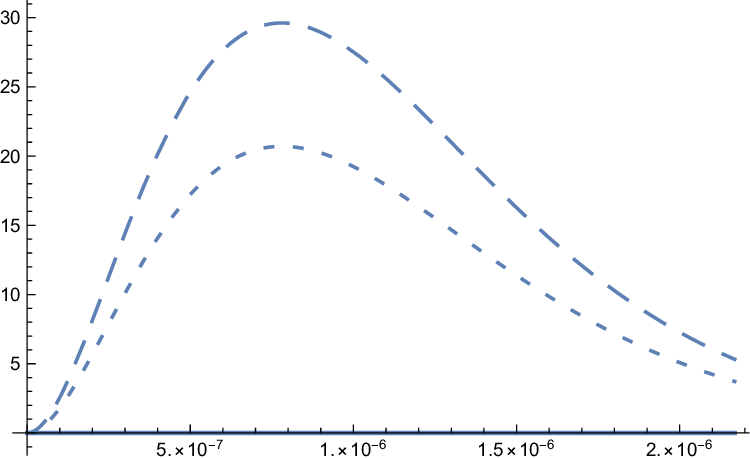}\\
		\hspace{-3.0cm}$r=\frac{\Delta}{m_a}-1 \rightarrow $  \\
		\caption{The same as in Fig. \ref{fig:pSP} but for the oxygen target. The $A$ term has been suppressed by a factor of 1/500, i.e. its actual value is 500 larger. Now $m_i,\,i=A,D$ with
			%	obtained from $r$ by $\Delta=m_i (1+r),\, i=A,D$ with the parameters $m_i$ given by
			$m_A=\frac{5}{3}\delta$, $M_D=\epsilon+\frac{5}{3}\delta$,  $\epsilon=0.0197$  eV. The other two transition types $ B,\,C $ do not occur.
		 The other two transition types $ B,\,C $ do not occur.
		}
		\label{fig:oxygen}
	\end{center}
\end{figure}
\begin{figure}
	\begin{center}
		\rotatebox{90}{\hspace{-0.0cm} $R \rightarrow $ per mol-year}
		\includegraphics[width=0.5\textwidth]{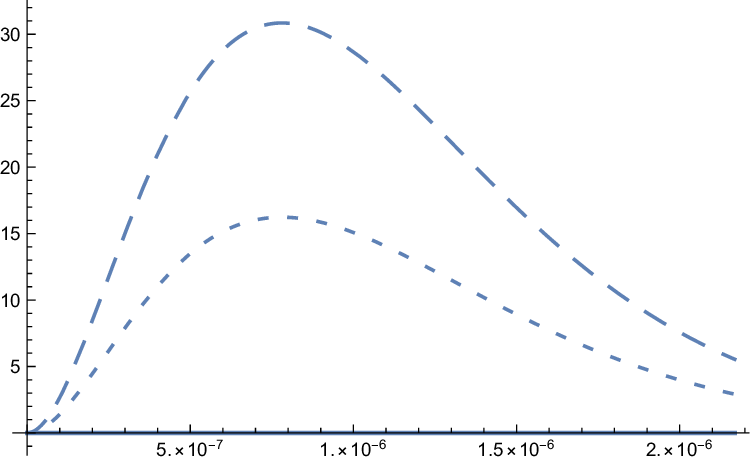}\\
		\hspace{-3.0cm}$r=\frac{\Delta}{m_a}-1 \rightarrow $  \\
		\caption{The same as in Fig. \ref{fig:pSP} but for the iron target. The $A$ term has been suppressed by a factor of 1/500, i.e. its actual value is 500 larger. Now $m_i,\,i=A,D$ with
		%	obtained from $r$ by $\Delta=m_i (1+r),\, i=A,D$ with the parameters $m_i$ given by
			$m_A=\frac{16}{5}\delta$, $M_D=\epsilon+\frac{16}{5}\delta$, $\epsilon=0.05$  eV. The other two transition types $ B,\,C $ do not occur.
		}
		\label{fig:Fe26}
	\end{center}
\end{figure}

 
 
ii) The second  interesting example of many particle configurations is that  of $_{26}$Fe allowing the transition $^5D_4\rightarrow ^5D_3$. \\The obtained results  presented in fig. 	\ref{fig:Fe26}.

At this point we should mention that, in the case of configurations with more than a single electron, some of the rates are enhanced due to the large spin interaction involved, see table 	\ref{tab:tab2}.
\subsection{A summary of the obtained results}
We have exhibited the results for the axion induced excitations for a number of atomic targets, see Figs \ref{fig:sSP} - \ref{fig:Fe26}. Similar results are expected for targets involving different radial functions with the same angular momentum structure as, e.g.,  in  single particle structures  involving s, p, and  d orbitals, Si instead of C, Zr instead Ti etc. The  expected event rates, the width of the resonance and the dependence on the magnetic field will be the same. Only the extracted value of the axion mass from the $B$, $C$ and $D$ terms will be different, since it depends on the experimentally determined spin orbit splitting $\epsilon$. 
 
We have seen that in all atoms considered the expected resonances are very narrow and, since we have no experimental or theoretical information about the axion mass, they might be missed by experiments. The location of the resonance, however, depends on the magnetic field employed. One, thus, may consider  a  magnetic field whose magnitude is changing periodically 
%in suitable steps 
from a minimum to a maximum value many times during the experiment. The oscillation period must be as short as possible and, in any case, very much shorter than the run time of the experiment. The latter must be  longer than the time implied by the predicted event rate to compensate for the fact that only a fraction of the time the equipment is going to be at the right state of sensitivity for axion detection.
% During each step the magnetic field remains constant. 
%The duration of each step can be selected to be short, provided that it is long enough so that 
%the atom reaches equilibrium after each step.
%is in but is should step s field with a period a very small fraction of the duration of the experiment. The latter maybe long, of the order of a year, to allow the accumulation of a sufficient number of events.
With this arrangement one can perhaps see the axion provided its mass lies between $m_i|_{B=B_{min}}$  and $m_i|_{B=B_{max}}$ for $i=A,B,C,D$. This range depends on the atom considered. The most favored case is the one with as large as possible  magnetic moment splitting. Thus in the case of the iron target one can detect light axion masses through the $A$ term in the range of   $\frac{16}{5}(\delta/n)\le m_a\le \frac{16}{5}\delta$. In particular for n=100 and $B=1$T one can detect axions 
in the mass range $2  \times 10^{-6}\mbox{eV}\le m_a\le 2 \times 10^{-4}\mbox{eV}$. This range is 
a bit wider than that of the  dedicated experiments involving resonance cavities, such as the well known ADMX and ADMX-HF and CAP \cite{CAPP}, searching for axion masses , see, e.g., \cite{Bibber16},\cite{PriSecSad88}, \cite{Stern14}, a summary \cite{MultIBSTh} and a recent review \cite{Ringwald16} giving the range $10^{-6}\mbox{eV}\le m_a\le10^{-4}\mbox{eV}$ (values $ m_a\le 4.4\mu$eV have recently been excluded by ADMX \cite{ADMX21}).\\
%\cite{ExpSetUp11b},\cite{ADMX10},\cite{Stern14,MultIBSExp} and CAPP  \cite{CAPP}, \cite{ExpSetUp11a}, \cite{Semer20a},\cite{Semer20b},\cite{SemerArch19}. 
Similarly for heavier axions through the $D$ term  one gets  a relation depending on the spin orbit splitting, in this case $0.05+2  \times 10^{-6}\mbox{eV}\le m_a\le 0.05 +2\times 10^{-4}\mbox{eV}$.

The width of the window, of course, can increase, if larger   magnetic fields are employed, and the minimum can be selected as convenient. 
\\
 Anyway such windows of axion mass can be open in  atomic physics detection, provided that the
sweep in frequency is smooth and gap-less, exploiting  
%More specifically such an experiment can benefit from 
% this connection we mention that the 
%experiments like those  proposed in
%this paper can benefit from 
% from  spectroscopic studies devoted to atoms in homogeneous magnetic fields,
%where constant "scaled energy" spectra were recorded. The scaled energy $\tilde{\epsilon} = E/B^{2/3}$, where
%E is the excitation energy and B the magnetic field strength (both in atomic units). So, in order to have
%a constant  $\tilde{\epsilon}$ 
the feasibility of simultaneous scan of the frequency and the magnetic field
%.during a scan, the laser frequency and the magnetic field had to vary simultaneously,
in a prescribed way (see, e.g., \cite{DTHVH94}, \cite{Leach10}). 
%The same can be done for the experiments proposed in
%the paper. That is the light frequency and the magnetic field can be scanned simultaneously
%in order for the laser to be always on resonance.

This way  one would think that the narrowness
of the signal is beneficial rather than problematic, yielding an advantage of the atomic experiments.
%, with the photons produced either by the de-excitation of the state populated by the axion or by the de-excitation of a suitably chosen 3nd level as proposed by Sikivie.

\section{Conclusions}

\label{conclusions}
In this paper we considered the possibility of direct detection of axion as a  dark matter candidate by measuring the rates for axion induced  atomic excitations. The essential input in our calculations was  strength of the  axion electron interaction  $g_{ae}/f_a$ and the axion flux on the detector. For the latter  we have used the standard halo parameters with a Maxwell-Boltzan distribution transformed in the local frame. The strength of the interaction    $g_{ae}/f_a$ was assumed to be  equal to  the limit obtained from the Borexino experiment. This assumption   allows  axion masses in the range that can be exploited by spin induced atomic transitions. That is  tens of $\mu$eV  within members of the same  multiplet, i.e.   $|J_1,M_1=-J_1\rangle\rightarrow |J_1,M_1=-J+1\rangle, J_1\ne0$ of the type $A$, and  axion masses in the range 1meV-1eV involving transitions of to the type  $|J_1,M=-J_1\rangle\rightarrow |J_2,M_2=-J_1+q\rangle,q=-1,0,1$, of the type $B$, $C$ and $D$, allowed by the angular momentum selection rules.

Furthermore, since the axion is absorbed by the atom, the calculated cross section exhibits resonance behavior. The resulting  pattern reflects the parameters of the  
% The location of the resonance as well as its width  are proportional to the escape velocity of
 velocity  distribution in the local frame and the momentum dependence of the axion electron interaction.
 The obtained  results  depend, of course, on the atom considered, through the parameters $\epsilon$ (the spin orbit splitting) as well as the $\delta$ ( the energy  splitting due to the magnetic moment interaction). The last two parameters  determine the  axion mass that can be detected, which is very close to the excitation energy. In addition the the resonance  behavior can  be exploited by experiments in minimizing   any background events.

In the special case of the type $A$ transitions the obtained rates, as shown at  the top of the resonance, are quite large as a result of the large axion flux. The highest rates obtained occur in the case of light axions. They involve  all the  $s_{1/2}$ transitions in any atom and the A type transitions  and for  the many   electron configuration atoms like Oxygen and $_{28}$Fe, namely $R=7.0\times  10^3$, $R=1.4\times  10^4$ and $R=1.4\times  10^4$ per mole-y respectively. 
%Rates for heavier axions (B,C,D transitions)  are expected in the range of (0.5-5.0)$\times 10^2$ per mole-y .
 The experimental detection difficulty  in this case is not connected with the expected rate but, as explained in section  \ref{ec:lowtemp},  with the very low temperature behavior of the target. The excited state must be essentially empty  of electrons. Furthermore  the target must exhibit atomic behavior at such low temperatures. It may be an advantage that such high  rates allow one to consider the  atom of interest in  the form of an impurity, at the level $1/10^3$ or even $1/10^4$ in an otherwise inert target. 

For the $B$, $C$,and $D$ type  transitions the low temperature requirements are not very stringent. The expected rates, as appearing at  the tops of  some of the the $B$, $C$, $D$ patterns in the  figures,  are much smaller than those of the $A$ terms, see Figs \ref{fig:pSP}-\ref{fig:Fe26}, but perhaps detectable. In the case of the    $_{22}$Ti a rate as   high   as $R=5.0\times  10^2$ per mole-y is obtained.
% Detectable rates for some of the other  targets considered in this work are expected.

The main experimental problem for  all types of transitions is due to  the fact that all resonances are very narrow and they might   be missed by the experiments. This is, unfortunately, so  since there is no  experimental or theoretical guide about the expected value of the axion mass. We have seen, however, that, if a suitable periodic  magnetic field is selected, whose magnitude  is in a  appropriate range  and its period is very much smaller than  the experimental run time, a window of axion mass, in the range of a  fraction of an meV wide, becomes open, which  may be adequate.

% Using reasonable values for the input parameters, like the  axion mass, the axion-electron coupling $g_e$ as well as the axion elctron interaction coefficients given in tables 	\ref{tab:tab1}-\ref{tab:tab2} and  the standard halo model,  we  get an expression for the  rate provided  by Eq (\ref{Eq:rate}).\\ Using this expression for axions in tens of $\mu$eV we get a detectable rate using a model value for the unknown parameter $g^2_e\approx 1/36$.\\ For axion mass in the meV-eV range the similarly  predicted rates become huge. If axions in this mass range exist, they can be detectable  even if  $g^2_e$ happens to be  $10^{-4}-10^{-5}$ smaller than above,  depending on the atom considered.\\ 
%$\left (C_{j_1,m_1,j_2,m_2,\ell}\right)^2 $. \\
%This result must be multiplied by the geometric factors, once a target is selected, which have been tabulated.
 % Furthermore the he rate exhibits a resonance behavior as shown in Figs  \ref{fig:axionefac} and \ref{fig:axionefac}. The  width of the resonance depends, among other things, the magnetic quantum numbers involved and the size of the axion mass.
%   \\ The experiments must be performed at low temperatures. We have seen that a very low temperature may be required for light axion. We hope that the atoms considered here may  suggest some possibilities of materials exhibiting atomic structure at sufficiently low temperatures. 

\section*{Acknowledgments} 
J.D.V  is happy to acknowledge that this work was supported by  IBS-R017-D1-2020-a00. Special thanks to professor  Yannis Semertzidis, director of the Center for Axion and Precision  Physics  Research, IBS,  at KAIST University, for his hospitality,  encouragement and  useful discussions. All authors are indebted to Professor P. Sikivie for a careful reading  of the manuscript and his very useful comments and suggestions as well as bringing to their attention the need of simultaneous scan of frequency and magnetic field. Special thanks to Professor H. Ejiri for clarifying some aspects of the experiment proposed in this work.

%\bibliography{TeX}

\section{Appendix: The modulation of the widths}
\label{sec:Appendix}
\begin{figure}
	\begin{center}
		\rotatebox{90}{\hspace{-0.0cm} $\frac{\Gamma_{mod}}{\Gamma_{av}}\rightarrow$}
		\includegraphics[width=0.7\textwidth,height=0.4\textwidth]{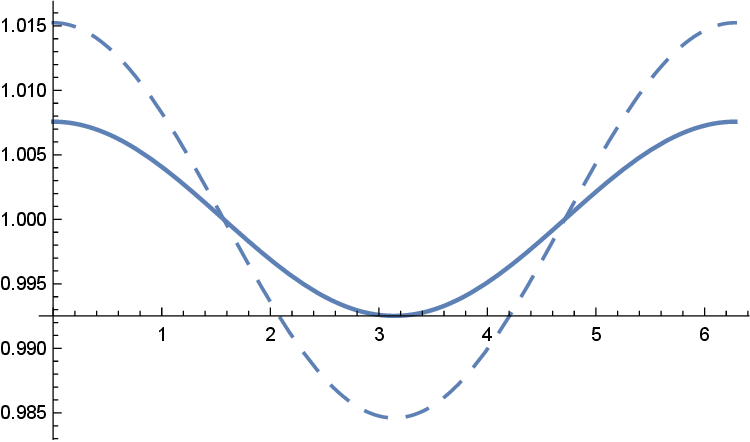}\\
		\hspace{-3.0cm}$\alpha \rightarrow$  \\
		\rotatebox{90}{\hspace{-0.0cm} $\frac{\Gamma_{mod}}{\Gamma_{av}}\rightarrow$}
		\includegraphics[width=0.7\textwidth,height=0.4\textwidth]{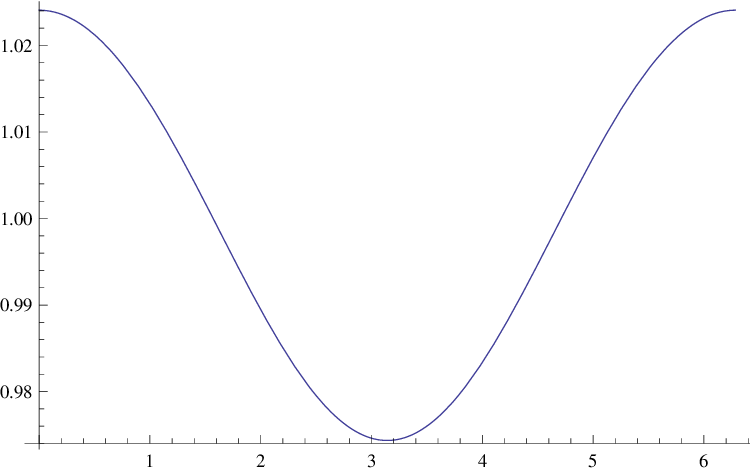}\\
		\hspace{-3.0cm}$\alpha \rightarrow$  \\
		\caption{In the top panel we exhibit the modulation of the width $\Gamma$, relative to its average value, as a function of the phase of the Earth. The notation for the curves is the same as in Fig. \ref{fig:velfac}.
			For comparison we present in the bottom panel the modulation curve obtained in the case of the standard axion to photon conversion, obtained with the same halo parameters \cite{VerSem16}.
		}
		\label{fig:fmod}
	\end{center}
\end{figure}

\begin{figure}
	\begin{center}
		\rotatebox{90}{\hspace{-0.0cm} $\sigma \rightarrow$}
		\includegraphics[width=0.7\textwidth,height=0.4\textwidth]{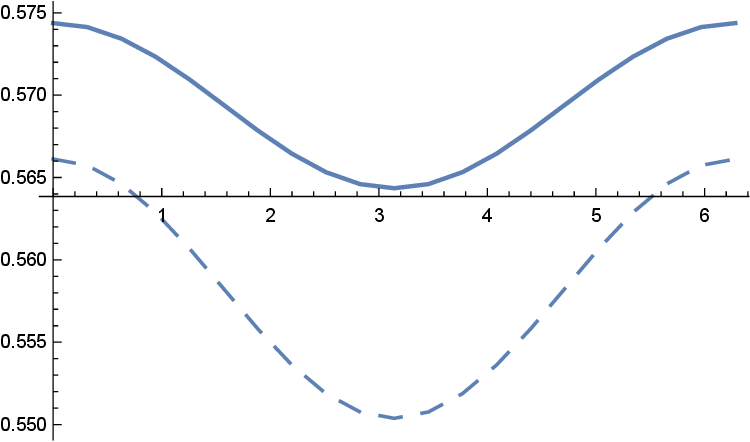}\\
		\hspace{-3.0cm}$\alpha \rightarrow$  \\
		\caption{The time variation of the width for axion absorption by an atom due to the motion of the Earth. The notation for the curves is the same as in Fig. \ref{fig:fmod}.
		}
		\label{fig:dispersion}
	\end{center}
\end{figure}

   The modulation of the widths  can be simply included by making in the local frame we make the replacement:
\beq
F_0(X)\rightarrow  e^{-\delta \cos{\alpha}-\delta ^2} F_0\left (X \left(\frac{1}{2} \delta  \cos \alpha +1\right)\right ),
\eeq
where $\alpha$ is the phase of the Earth, $\alpha=0$ around June 3nd, and $\delta$ the ratio of Earth's  velocity around the sun divided by the sun's velocity around the galaxy,  $\delta\approx 0.135$. We thus get a time variation of the width shown in Fig. \ref{fig:fmod}. We see that the effect is small, the difference between the maximum and the minimum is less than $3\%$, almost the same with that obtained in the axion to photon conversion \cite{VerSem16}. We note, however that, in addition to the seasonal dependence we have a  dependence on the magnetic quantum numbers of the states involved The variation in the case of $m_1\ne m_2$ is almost twice as large compared to that with $m_1 = m_2$.
%	This, in principle, can be exploited by the experiments.
\\It is amusing to know that the dispersion $\sigma=\sqrt{\langle X^2 \rangle -\langle X\rangle^2}$ also exhibits a time dependence (see Fig.  \ref{fig:dispersion}).

\end{document}